\documentclass[11pt,a4paper]{article}
\usepackage{jheppub}
\usepackage[utf8]{inputenc}
\usepackage{graphicx,fancybox,float,comment}
\usepackage{pdflscape}
\usepackage{units}
\usepackage{multirow,array,arydshln}
\usepackage[dvipsnames]{xcolor}
\usepackage{braket,tensor}
\usepackage{amsmath,amssymb,amsfonts,mathrsfs}
\usepackage{tikz-cd}
\usepackage{mdframed}
\usepackage{subfig}

\usepackage{ulem}

\global\interfootnotelinepenalty=1000

\graphicspath{{graphics}}

\newcommand{\bea}{\begin{eqnarray}}
\newcommand{\eea}{\end{eqnarray}}
\newcommand{\be}{\begin{equation}}
\newcommand{\ee}{\end{equation}}

\newcommand{\lr}[1]{\left( #1 \right)}

\def\({\left(}
\def\){\right)}

\def \a {\alpha}

\def \d {\delta}

\def \r {\rho}
\def \l {\lambda}
\def \m {\mu}
\def \n {\nu}
\def \s {\sigma}

\def\nn {\nonumber}

\subheader{\begin{flushright}
\end{flushright}}

\title{Higher derivative holography and temperature dependence of QGP viscosities}

\author[]{Thomas Apostolidis$^{a}$,}
\author[]{Umut G\"ursoy$^{b}$,}
\author[]{Edwan Pr\'eau$^{b}$}
\affiliation[a]{ Universit\'e Paris Cit\'e, Astroparticule et Cosmologie, F-75006 Paris,
France\\}
\affiliation[b]{Institute for Theoretical Physics and Center for Extreme Matter and Emergent Phenomena,\\ Utrecht University, 3584 CC Utrecht, The Netherlands}
\emailAdd{apostolidis@apc.in2p3.fr}
\emailAdd{u.gursoy@uu.nl}
\emailAdd{e.c.m.preau@uu.nl}

\abstract{Recent Bayesian analyses of heavy ion collision data have established a non-trivial temperature dependence of the shear and bulk viscosity per entropy. Motivated by this, we consider higher derivative corrections to realistic, bottom-up holographic models of quark-gluon plasma based on five-dimensional Einstein-dilaton theories and determine the dilaton potentials in the higher derivative terms by matching the Bayesian analyses. A byproduct of our analysis is the bulk viscosity that follows from the holographic V-QCD theory. Higher derivative corrections when treated perturbatively lead to tension with existing data. We investigate possible resolutions.\\
\begin{center}
\textit{Dedicated to the memory of Umut G\"{u}rsoy, who initiated this project and took part in every stage of its development}.
\end{center}}

\begin{document}
\maketitle
\flushbottom

\section{Introduction}
\label{sec::intro}

Heavy-ion collisions at RHIC and LHC provide a unique window into the transport properties of QCD; see e.g. \cite{Busza:2018rrf} for a recent review. The transport properties of the quark-gluon plasma produced in these experiments can be characterized by the transport coefficients that enter the hydrodynamic description of the fluid, most prominent of which are the shear and the bulk viscosities that are usually defined per entropy, i.e. $\eta/s$ and $\zeta/s$ respectively\footnote{Another coefficient at first-order in the hydrodynamic derivative expansion, conductivity, is relevant at finite quark density or in the presence of electromagnetic fields, which are ignored here.}. Knowledge of these coefficients requires computing stress tensor retarded Green's functions at finite temperature. The only available first-principles method, lattice QCD, is not quite suitable for this task: it is based on the Euclidean formulation, and the analytic continuation to the Lorentzian signature contains large systematic errors, as this requires knowledge of the entire particle spectrum of QCD; see \cite{Meyer:2007ic,Meyer:2007dy}.  

Recently, a more phenomenological approach, based on Bayesian analysis of heavy-ion data, became available \cite{Nijs:2020ors,Nijs:2020roc,Nijs:2023yab,Giacalone:2023cet,JETSCAPE:2020mzn}, see section \ref{sec::Bayes}. Here, one models the evolution of the quark-gluon plasma by hydrodynamics, e.g. 14-moment M\"uller-Israel-Stewart model \cite{Denicol:2014vaa}, supplemented by initial conditions for the energy-momentum distribution of nucleons,  e.g. provided by the Trento model \cite{Moreland:2014oya} and a standard freezout procedure \cite{Cooper:1974mv} to obtain the hadron distributions that are subsequently matched onto experimental data. This provides a probability distribution for the transport coefficients that characterize the hydrodynamic model. This approach also contains significant systematic errors resulting mainly from the modeling of the initial conditions and hydrodynamic evolution and the choice of the data sets used for matching. This can be clearly seen by comparing Bayesian analyses of Jetscape \cite{JETSCAPE:2020mzn} versus Trajectum\footnote{The Trajectum results presented in this work come from the most recent analysis \cite{Giacalone:2023cet}, which presents important differences with the original calculation in \cite{Nijs:2020ors,Nijs:2020roc}. Even though the results for the viscosities were not presented in \cite{Giacalone:2023cet}, they were actually also accessible from the numerical results. The model used for the Bayesian analysis was the same as in \cite{Nijs:2023yab}, which contains more details. We thank Wilke van der Schee and Govert Nijs for sharing the relevant data with us.} \cite{Nijs:2020ors,Nijs:2020roc,Nijs:2023yab,Giacalone:2023cet}; see Figures \ref{exp visc figure} and \ref{bulk visc figure}. A crucial aspect that resulted from these analyses is the non-trivial temperature dependence\footnote{See e.g. \cite{Trachenko:2020ktm} and \cite{Cremonini:2012ny} for a discussion of the temperature dependence of $\eta/s$ in the QGP, including a comparison with other fluids in nature.} of $\eta/s$ and $\zeta/s$. 

Yet another complementary tool is provided by the holographic correspondence \cite{Maldacena:1997re,Gubser:1998bc,Witten:1998qj}. A triumph of this approach was the calculation of $\eta/s$ \cite{Policastro:2001yc} at strong coupling that resulted in the universal value $1/(4\pi)$ that agrees well with the aforementioned Bayesian analyses \cite{Nijs:2020ors,Nijs:2020roc,Nijs:2023yab,Giacalone:2023cet,JETSCAPE:2020mzn}. Five-dimensional ``bottom-up" holographic models of QCD-like large-N gauge theories have been developed to include the salient features of the theory such as confinement, gapped hadron spectrum, chiral symmetry breaking, QCD anomalies, etc. The type of models we will consider in this work are the large N improved holographic QCD (ihQCD) model \cite{Gursoy:2007cb,Gursoy:2007er} and its extension to include fundamental matter \cite{Jarvinen:2011qe} in the Veneziano limit \cite{Veneziano:1976wm}, with large N and large number of flavors with their ratio fixed. These theories naturally incorporate the running of the 't Hooft coupling through a non-trivial profile of the dilaton $\Phi$ in the holographic dimension. The latter is prompted by introduction of a dilaton potential $V(\Phi)$, whose large $\Phi$ asymptotics (corresponding to the IR of the QFT, and the deep interior of the bulk geometry) are judiciously chosen such that the theory is confining. The extension of ihQCD and V-QCD to finite temperature was studied in \cite{Gursoy:2008bu,Gursoy:2008za,Gursoy:2009jd} and \cite{Alho:2012mh,Alho:2013hsa} respectively, and incorporation of electromagnetic fields --- relevant for off-central heavy-ion collisions and neutron star mergers --- was considered in \cite{Gursoy:2016ofp,Gursoy:2017wzz}. 

All these holographic theories are built on two-derivative Einstein gravity, which universally fixes the value of $\eta/s = 1/(4\pi)$ \cite{Buchel:2003tz}. In particular, the aforementioned non-trivial temperature dependence observed in experiment requires inclusion of $1/N$ or higher derivative corrections, realized, respectively, as loop and $\alpha'$ contributions in string theory. On the other hand, the temperature dependence of $\zeta/s$ is already present at the two-derivative level and was investigated in the holographic models of QCD-like theories in \cite{Gubser:2008sz,Gursoy:2009kk}. See \cite{Buchel:2007mf,Gubser:2008sz,Buchel:2008uu,Gursoy:2009kk,Eling:2011ms,Buchel:2011wx,Buchel:2011uj,Ballon-Bayona:2021tzw,Demircik:2023lsn,Demircik:2024bxd,Buchel:2023fst,CruzRojas:2024etx} for a non-exhaustive list of other literature on bulk viscosity in the context of holography. Hence, one needs to incorporate the higher derivative corrections to Einstein-dilaton type holographic models to capture the nontrivial temperature profile of both $\eta/s$ and $\zeta/s$, as first observed in \cite{Cremonini:2012ny}. 

Such corrections have been considered early on in the context of violations of the conjectured KSS bound $\eta/s \geq 1/(4\pi)$ \cite{Kovtun:2003wp,Kovtun:2004de}; see e.g. \cite{Kats:2007mq,Brigante:2007nu,Myers:2008yi} and \cite{Cremonini:2011iq} for a review. In \cite{Grozdanov:2016zjj,Folkestad:2019lam}, their effect was also studied in a dynamical holographic toy model for heavy-ion collisions. However, these investigations were concerned with holographic duals of conformal field theories which are not suitable for QCD. On the other hand, literature on stringy corrections to holographic theories of QCD-like non-conformal and confining theories is rather limited\footnote{See \cite{Buchel:2010wf} for the first such example. Also note the references \cite{Czajka:2018bod,Yadav:2020tyo,Kushwah:2024ngr}, which investigated both $\a'$ and some type of string loop corrections in a top-down model.}. Stringy corrections to ihQCD are discussed in \cite{Gursoy:2007cb,Kiritsis:2009hu,Alho:2015zua} but not in the context of transport. Higher curvature corrections to shear transport in duals of Einstein-dilaton theories were studied in detail in \cite{Cremonini:2012ny}, but no reliable temperature profile was available from heavy-ion data back then for neither $\eta/s$ nor $\zeta/s$, so these corrections were not sufficiently constrained. Recently, there has been a revival of interest in determining the transport coefficients in holographic theories, including higher curvature corrections. While \cite{Demircik:2023lsn,Demircik:2024bxd} introduced a novel method to compute all transport coefficients by varying the background solutions, which in principle can be extended directly to higher curvature theories \cite{Gallegos}, Buchel, Cremonini and Early \cite{Buchel:2023fst} extended the standard techniques to beyond the supergravity approximation in generic theories, which is precisely the problem we want to address in this paper. 

The question we start with is whether we can construct realistic (in the aforementioned sense) holographic theories for the quark-gluon plasma beyond the supergravity approximation, which postdict the non-trivial temperature profile of $\eta/s$ and $\zeta/s$ observed in the Bayesian analyses \cite{Nijs:2020ors,Nijs:2020roc,Nijs:2023yab,Giacalone:2023cet,JETSCAPE:2020mzn}. This is a sensible approach to characterizing transport in strongly interacting quark-gluon matter, as it attempts to merge the two existing approaches, Bayesian analysis of heavy-ion data and holography, in the absence of a first-principle derivation.

Specifically, we consider a 5D Einstein-dilaton theory with a dilaton potential $V(\Phi)$ and with fourth-order derivative corrections. There are three types of corrections at this order, given by $R^2$, $R_{\mu\nu}^2$ and $R_{\mu\nu\alpha\beta}^2$. We derive the general expressions for the viscosities in presence of such corrections\footnote{Only the bulk viscosity is a new result, since the shear viscosity was computed in \cite{Cremonini:2012ny}.}, where we allow for dilaton dependence of the coefficients $G_i(\Phi)$ in order to have sufficient freedom to match the temperature profiles of $\eta/s$ and $\zeta/s$. We then analyze the case of a single Riemann-squared correction with coefficient $G(\Phi)$, and investigate numerically the possibility to match the Bayesian results in this case. Note that, as shown in \cite{Kats:2007mq} and reviewed in Appendix \ref{sec::appC}, this is the most general type of correction for shear viscocity (but not for bulk viscosity). 

Our main results are: 
\begin{enumerate}
\item The potential $G(\Phi)$ can be constructed to fit the Trajectum result for $\eta/s$ \cite{Nijs:2020ors,Nijs:2020roc,Nijs:2023yab,Giacalone:2023cet}, using the analytic holographic expression for the shear viscosity per entropy derived in \cite{Cremonini:2012ny}. However, this implies that the corresponding corrections on the background would have to be large, thus invalidating our perturbative framework for the curvature corrections. This implies that the finite-coupling corrections are indispensable for transport in the quark-gluon plasma. 
\item Based on Trajcetum results, we investigated the influence of a (very mild) decrease with temperature in $\eta/s$ on the bulk viscosity. Our results indicate that $\zeta/s$ tends to increase at low temperature, while it becomes smaller at high temperature, hence moving further away from the Trajectum band. This suggests that other types of correction are necessary to fit both $\eta/s$ and $\zeta/s$ with reasonable accuracy. 
\item Before even adding the curvature corrections we calculate the zeroth order result for $\zeta/s$ in the V-QCD model for the first time. Our findings hint towards a sizable breaking of conformality in the quark-gluon plasma.

\end{enumerate}

We organized the rest of the paper as follows. In the next section, we summarize the relevant results from existing Bayesian analysis. In Section \ref{sec::viscos} we outline the holographic calculation of shear and bulk viscosity and the holographic models that we employ to fit the Bayesian data, with details in Appendices \ref{sec::bg}, \ref{sec::appB} and \ref{sec::appC}. Section \ref{sec::fit} is devoted to fitting the holographic models to results of Bayesian analyses. We discuss our results and provide an outlook to future work in section \ref{sec:dis}. Several appendices \ref{sec::bg}-\ref{sec::appE} provide details of our calculations but also contain complementary information. In Appendix \ref{sec::bg}, we analyze the background equations including the higher derivative corrections and work out the near boundary and deep interior limits of the background. In particular we obtain the asymptotic behavior of the $G$ function such that it does not spoil the salient features, e.g. confinement, of our holographic QCD models. This is important also to justify our omission of curvature corrections to the background. For completeness, in Appendices \ref{sec::appB}, \ref{sec::appD} and \ref{sec::appE} we recapitulate the calculations of shear and bulk viscosities including higher derivative corrections. In Appendix \ref{sec::appC} we review the result of \cite{Kats:2007mq} that only the Riemann squared correction contributes to $\eta/s$.

\section{Bayesian analysis of Heavy Ion Collisions}
\label{sec::Bayes}

In complex systems such as the quark-gluon plasma, the correlation between microscopic parameters of the theory, e.g. the shear and bulk viscosity per entropy, and observables, e.g. the distribution of soft hadrons observed at the detector, is complicated. As a result, inferring the values of the microscopic parameters from observed data, i.e. the ``inverse problem," becomes extremely difficult. A systematic means to address this problem is provided by the Bayes' theorem
\be
{\cal P}(B|A) = \frac{{\cal P}(A|B) {\cal P}(B)}{{\cal P}(A)}\, ,
\ee
where ${\cal P}(B|A)$ denotes the conditional probability that $B$ occurs given condition $A$. For example, if $A$ is a collection of observables in collisions and $B$ is a set of microscopic variables, we can construct ${\cal P}(A|B)$ by running many simulations of the collisions based on the hydrodynamic model with parameters $B$. Bayes' theorem then provides a probability distribution for the microscopic parameters given the actual observed data.    

The Bayesian method is flawed by an inherent ambiguity, originating from the choice of parametrization for the microscopic parameters and the selection of observables. It would be surprising if quantitatively similar results were obtained for different choices of parametrization. Indeed, the two existing analyses \cite{JETSCAPE:2020mzn} and \cite{Nijs:2023yab,Giacalone:2023cet} differ both in the parametrization of temperature dependence for the viscosities and in the selected observables. In particular, 
in \cite{Nijs:2023yab,Giacalone:2023cet} $\eta/s$ is parametrized as a piecewise linear function of the temperature: constant up to $0.15\,\text{GeV}$, with slope $(\eta/s)_{\text{slope}}$ and average $\overline{\eta/s}$ for $0.15\,\text{GeV}<T<0.3\,\text{GeV}$, then slope $(\eta/s)_{\text{slope}}+(\eta/s)_{\d \text{slope}}$ up to $0.5\,\text{GeV}$, and reaching some value $(\eta/s)_{0.8\,\text{GeV}}$ at $T=0.8\,\text{GeV}$ and beyond.
$\overline{\eta/s},(\eta/s)_\text{slope},(\eta/s)_{\d\text{slope}}$ and $(\eta/s)_{\text{0.8\,\text{GeV}}}$ are the parameters of the model, to be determined by the Bayesian analysis. 

On the other hand, in \cite{JETSCAPE:2020mzn} the parametrization is of the form\footnote{Note that this choice does not rely on any specific model. It is just a simple ansatz that can accommodate the expected feature of a minimum in $\eta/s$ around deconfinement \cite{Cremonini:2012ny,Trachenko:2020ktm}. This minimum is assumed here to lie precisely at the cross-over temperature $T_c$.}
\begin{equation}
\label{exp eta over s J}
    \frac{\eta}{s} = a_{\textrm{low}}(T-T_c)\Theta(T_c - T) + \eta_0 + a_{\textrm{high}}(T_c - T)\Theta(T-T_c)\, ,
\end{equation}
where $T_c=154\times 10^{-3}\,\text{GeV}$ is the deconfinement cross-over temperature, and the parameters $a_{\textrm{low}}$, $a_{\textrm{high}}$ and the kink $\eta_0$ are to be determined. Bulk viscosity per entropy is parametrized as a skewed Cauchy distribution in both works
\begin{equation}
    \frac{\zeta}{s}=d\left[\frac{e^2}{\lr{T-T_\zeta}^2+e^2}\right]\, .
\end{equation}
While the constants $d$, $e$ and $T_\zeta$ are left as parameters to be determined in \cite{Nijs:2023yab,Giacalone:2023cet}, \cite{JETSCAPE:2020mzn} further parametrizes $e = w_\zeta (1+\lambda_\zeta \textrm{sgn}(T-T_\zeta))$. 
As mentioned above, not only the parametrization of transport coefficients, but also the choice of observables differ in the two works. Jetscape uses $p_T$-integrated flow observables and includes only heavy-ion collisions (lead-lead and gold-gold), whereas Trajectum works with $p_T$-differential flow observables and includes both lead-lead and proton-lead collisions. 

\begin{figure}[h]
    \centering
    \includegraphics[scale=1]{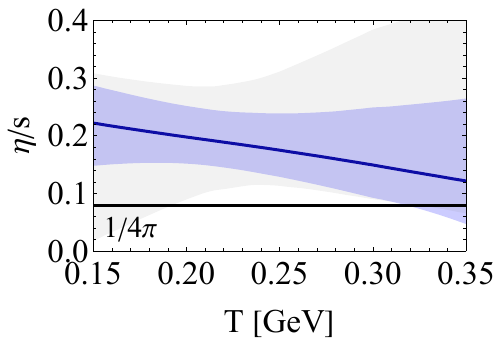}
    \includegraphics[]{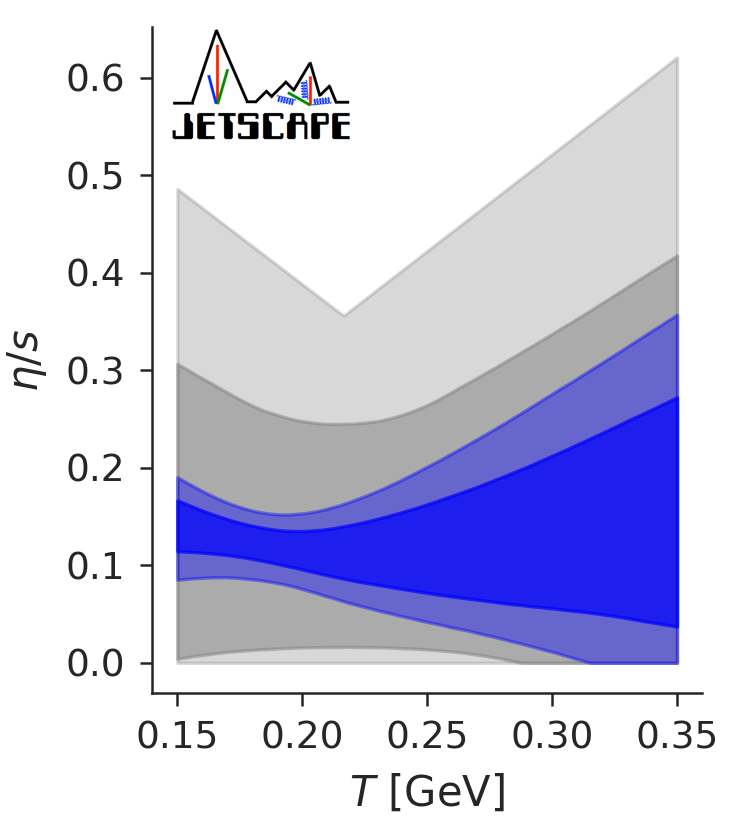}
    \caption{Posterior distribution for the shear
viscosity to entropy ratio versus temperature for Trajectum (left) and Jetscape (right). For Trajectum the 90$\%$ confidence band is shown in blue and the dark blue curve is the mean. The gray area signifies the 90$\%$ confidence band for the prior distribution. For Jetscape see Figure \ref{bulk visc figure} for definition of the confidence bands.}
    \label{exp visc figure}
\end{figure}

We present the most probable values of the parameters obtained in the Trajectum analysis \cite{Giacalone:2023cet} here for future reference: 
\begin{equation}
\nn
    \overline{\eta/s}=0.191^{+0.051}_{-0.048}\,, \quad (\eta/s)_\text{slope}=-0.34^{+0.66}_{-0.61}\,\text{GeV}^{-1}\,, \quad (\eta/s)_{\d\text{slope}}=-0.17^{+0.83}_{-0.80}\,\text{GeV}^{-1} \, ,
\end{equation}
\begin{equation}
\label{a,b,c with errors} (\eta/s)_{0.8\,\text{GeV}}=0.27^{+0.13}_{-0.14} \, ,
\end{equation}
\begin{equation}
\label{zeta errors}
    d=0.074^{+0.048}_{-0.044}\,, \quad  d\times e=0.0068^{+0.0033}_{-0.0029}\,\text{GeV}\,, \quad T_\zeta=0.330^{+0.098}_{-0.100}\,\text{GeV}\,.
\end{equation}
The resulting profiles for $\eta/s$ are shown in Figure \ref{exp visc figure}, along with the 90$\%$ confidence band.

\begin{figure}[h]
\centering
\begin{tabular}{ll}
\hspace{-0.5cm}
\includegraphics[scale=1]{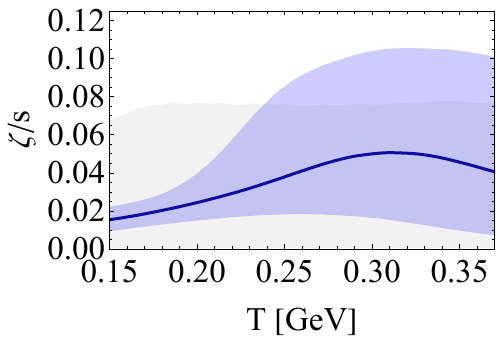}
&
\hspace{-0.3cm}\includegraphics[scale=0.75]{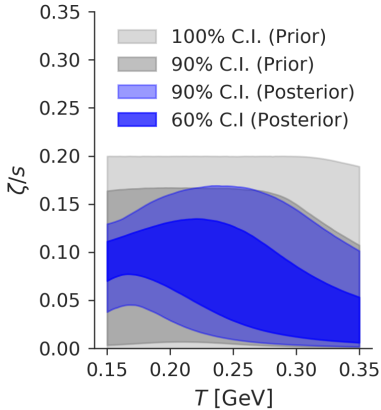}
\end{tabular}
\caption{Posterior distribution for the bulk
viscosity to entropy ratio versus temperature, with a
90$\%$ confidence band shown in blue, from Trajectum \cite{Giacalone:2023cet} (left) and Jetscape\cite{JETSCAPE:2020mzn} (right).}
\label{bulk visc figure}
\end{figure}

We also reproduce in Figure \ref{bulk visc figure} the bulk viscosity per entropy obtained in the two studies. Both results have a large uncertainty, and they are in statistical tension only at low temperatures, where Trajectum predicts smaller values of $\zeta/s$.  
\section{Holographic shear and bulk viscosities}\label{sec::viscos}

\subsection{Generalities}\label{sec::gen}
Kubo's linear response theory relates the shear and bulk viscosities to the stress tensor correlators. Shear deformation
corresponds to a traceless deformation of the metric which we can take on the xy plane with no loss of generality, using rotational symmetry: 
 \begin{equation}
    \eta=-\lim_{\omega\to 0} \frac{1}{\omega}\text{Im}G^R_{xy,xy}\lr{\omega,k=0}\, ,
\end{equation}
where the retarded Green's function is at vanishing momentum and in the thermal state. It is given by
\begin{equation}
    G_{x y, x y}^R(\omega, \mathbf{k}=0)=-i \int d t d \mathbf{x} e^{i \omega t} \theta(t)\left\langle\left[T_{x y}(x), T_{x y}(0)\right]\right\rangle \, .
\end{equation}
Bulk deformations correspond to a scaling of the spatial volume, which is related to the trace of the metric fluctuation. The corresponding Kubo formula reads \cite{Hosoya:1983id} 
\begin{equation}
    \zeta=-\frac{4}{9}\lim_{\omega\to 0}\frac{1}{\omega}\text{Im}G^R_{ii}{\omega}\, ,
\end{equation}
where the Green's function is given by 
\begin{equation}
    G^R_{ii}(\omega)=-i \int d t d \mathbf{x} e^{i \omega t} \theta(t)\left\langle\left[\frac{1}{2} T_i^i(t, \mathbf{x}), \frac{1}{2} T_j^j(0, \mathbf{0})\right]\right\rangle\,.
\end{equation}
In holography, the $\omega\to 0$ limit of the Green's functions follows from the corresponding fluctuation of the black-brane background, with non-normalizable boundary condition at the boundary and infalling boundary condition at the horizon\footnote{An alternative derivation involves formulating the fluctuation equations as background variations which has the advantage of providing analytic formulae for both shear and bulk viscosity \cite{Demircik:2023lsn,Demircik:2024bxd}.} \cite{Son:2002sd}.  

As mentioned in the introduction, we simplify our analysis by considering only the Riemann-square type corrections\footnote{It can actually be shown that the shear viscosity per entropy is not modified by the other types of curvature squared corrections $R^2$ and $R_{\m\n}^2$. The argument is presented in appendix \ref{sec::appC}.}. Hence the generic form of our action reads
\begin{equation}
\label{Riemann action}
    S=\frac{1}{16 \pi G_5} \int d^5 x \sqrt{-g}\left[R-\frac43(\nabla \Phi)^2+V(\Phi)+\ell^2 \beta G(\Phi) R_{\mu \nu \rho \sigma} R^{\mu \nu \rho \sigma}\right]\,,
\end{equation}
where $G_5$ is the 5D Newton constant, $\ell$ is the AdS length, $\beta$ is a parameter which controls the magnitude of the higher curvature corrections and $V$ and $G$ are generic functions of the dilaton $\Phi$. In particular, the background solution is pure $AdS_5$ for $V = 12/\ell^2$ and $G=0$. For the action \eqref{Riemann action}, the calculation of \cite{Cremonini:2012ny} implies that the shear viscosity per entropy\footnote{We note that the usual Bekenstein-Hawking formula for the entropy in terms of the area of the black brane horizon is modified to Wald's formula in the presence of higher curvature corrections \cite{Wald:1993nt}.} can be expressed completely in terms of the potentials in (\ref{Riemann action}) as 
\begin{equation}
\label{eta over s theory}
    \frac{\eta}{s}=\frac{1}{4\pi}\left[1+\frac{2}{3}\beta\ell^2\lr{-G\lr{\Phi_h}V\lr{\Phi_h}+\frac{9}{8}\partial_{\Phi}G\lr{\Phi_h}\partial_{\Phi}V\lr{\Phi_h}}\right]\, ,
\end{equation}
where $\Phi_h$ is the value of the dilaton at the horizon. A similar derivation\footnote{The details of the calculation can be found in appendix \ref{sec::appD}.} in the case of bulk viscosity per entropy (generalizing the result of \cite{Buchel:2023fst}) leads to 
\begin{equation}\label{zetas1}
    \frac{\zeta}{s}=\frac{8z_0^2}{27\pi} \left[1-\frac{1}{2} \beta  \ell ^2 \left(\partial_\Phi V(\Phi_h) \partial_\Phi G(\Phi_h)+\frac{(\partial_{\Phi}V(\Phi_h))^2 G(\Phi_h)}{V(\Phi_h)}-\frac{4}{3}  V(\Phi_h) G(\Phi_h) \right)\right] \, ,
\end{equation}
where $z_0$ is the horizon value of the gauge invariant dilaton fluctuation in the hydrodynamic limit (see appendix \ref{sec::appD}). The expression \eqref{zetas1} can readily be extended to the case where the action contains all three different types of 4-derivative corrections 
\bea\label{genac}
S =\frac{1}{16 \pi G_5} \int &&d^5 x \sqrt{-g}\bigg[R-\frac{4}{3}(\nabla \Phi)^2+V(\Phi)\nn\\ && +\ell^2 \beta\lr{G_1\lr{\Phi}R^2+G_2\lr{\Phi}R_{\mu\nu}R^{\mu\nu}+ G_3(\Phi) R_{\mu \nu \rho \sigma} R^{\mu \nu \rho \sigma}}\bigg]  \, .
\eea
The same analysis in this case yields the following bulk viscosity to entropy ratio (see appendix \ref{sec::appD}):
\begin{equation}\label{zetas2}
    \frac{\zeta}{s}=\frac{8z_0^2}{27\pi} \left[1-\frac{1}{2} \beta  \ell ^2 \left(\partial_\Phi V(\Phi_h) \partial_\Phi \tilde G(\Phi_h)+\frac{(\partial_{\Phi}V(\Phi_h))^2 \tilde G(\Phi_h)}{V(\Phi_h)}-\frac{4}{3}  V(\Phi_h) \tilde G(\Phi_h) \right)\right] \, ,
\end{equation}
where we defined $\tilde G \equiv 5G_1 + G_2 - G_3$. 

The formulae (\ref{eta over s theory}) and (\ref{zetas1}) will form the basis for our comparison with the Bayesian analysis below. In particular, \eqref{eta over s theory} can be seen as a differential equation for $G(\Phi_h)$, which implies that any given profile $\eta/s(T)$ can be reproduced\footnote{Note that going from $\eta/s(T)$ to $\eta/s(\Phi_h)$ also requires to know the temperature as a function of the horizon location $T=T(\Phi_h)$. The latter can be extracted in a standard way from the background solution, by computing the surface gravity at the horizon.} with the appropriate potential $G$. The explicit solution takes the form
\begin{equation}
\label{G solution}
G\lr{\Phi_h}=F^{-1}\lr{\Phi_h}\int d\Phi_h F\lr{\Phi_h}R\lr{\Phi_h} +c_1F^{-1}\lr{\Phi_h},
\end{equation}
where
\begin{equation}
F\lr{\Phi_h}\equiv\exp\left\{\int \left(-\frac{8}{9}\frac{V\lr{\Phi_h}}{\partial_{\Phi}V\lr{\Phi_h}}\right)d\Phi_h\right\} \, , \quad R\lr{\Phi_h}\equiv\left[4\pi\frac{\eta}{s}(T(\Phi_h))-1\right]\frac{4}{3\ell^2\beta \partial_{\Phi}V\lr{\Phi_h}} \, .
\end{equation}
The integration constant $c_1$ which multiplies the homogeneous solution in \eqref{G solution} does not affect the viscosity since it cancels in (\ref{eta over s theory}). 

\subsection{Holographic setup}\label{sec:setup}
As our starting point, we will consider the V-QCD model \cite{Jarvinen:2011qe}, which is itself an extension of improved holographic QCD (ihQCD) \cite{Gursoy:2007cb,Gursoy:2007er}. These models are based on the assumption that the operator product algebra of QCD  is semi-closed in the IR, and consists of only a small subset of UV marginal and irrelevant operators \cite{Shifman:1978bx,Novikov:1983jt,Novikov:1980uj}: the stress tensor $T_{\m\n}$, the scalar glueball operator $\textrm{tr} F^2$, chiral condensate $\bar q q$, left and right SU($N_f$) chiral currents $L^\mu$ and $R^\mu$, and the U(1)$_B$ baryon number current $V^\mu$\footnote{One can extend this to include the CP-odd operator $\textrm{tr} F \wedge F$ but this can be ignored for the purpose of calculating viscosities.}. The corresponding holographic bulk fields are the metric, dilaton, open string tachyon and gauge fields. These fields propagate on a five-dimensional asymptotically AdS background, which is a solution of an Einstein-dilaton theory with space-filling flavor branes. See \cite{Jarvinen:2021jbd}
for a review. A non-trivial dilaton potential $V(\Phi)$ yields a non-trivial profile for $\Phi$ in holographic coordinate $r$, which corresponds to the running of the QCD 't Hooft coupling $\lambda_{\text{QCD}}$ in the IR. Accordingly, the ``glue" part of the theory (pure Yang-Mills at large $N_c$), improved further by a Riemann-square correction, is given by the action 
\begin{equation}\label{gulac}
  S_{glue}=  M_p^{3}N_{c}^{2}\int d^{5}x\sqrt{-g}\left[R-\frac{4}{3}\lr{\partial\Phi}^2+V_g\lr{\Phi}+\ell^2 \beta G(\Phi) R_{\mu \nu \rho \sigma} R^{\mu \nu \rho \sigma}\right]\, ,
\end{equation}
where we wrote $16 \pi G_5 = 1/(M_p^3 N_c^2)$, with $N_c$ the number of colors and $M_p$ the 5D Planck mass, which controls the overall normalization of the free energy. A judicious choice of $V_g$ asymptotics for large and small $\lambda$ ensures confinement and asymptotic freedom in the dual theory \cite{Gursoy:2007cb,Gursoy:2007er}. Deconfinement is described by a Hawking-Page transition between the low temperature thermal graviton gas solution, and the high temperature black-brane solution \cite{Gursoy:2008za}. Thermodynamics in the deconfined black-brane phase agrees remarkably well with large $N_c$ extrapolations of lattice QCD simulations \cite{Panero:2009tv,Gursoy:2008bu,Gursoy:2009jd}.

V-QCD is obtained from ihQCD by coupling the ``flavor" part, which consists of a number $N_f$ of space-filling $D4$ and $\overline{D4}$ branes with the DBI type action 
\begin{equation}\label{flavac}
   S_{flavor}=-x_fM_p^3N_c^2\int d^5x V_{f0}\lr{\Phi}e^{-\tau^2}\sqrt{-\text{det}\lr{g_{\mu\nu}+\kappa\lr{\Phi}\partial_{\mu}\tau\partial_{\nu}\tau+w\lr{\Phi}\hat{F}_{\mu\nu}}}\, ,
\end{equation}
with
\begin{equation}
     x_f\equiv\frac{N_f}{N_c}\, , 
\end{equation}
a finite constant parameter\footnote{$x_f$ finite means that we consider the Veneziano large N limit \cite{Veneziano:1976wm} of the model.}.
The endpoints of open strings on the branes model left and right-handed quarks and the condensation of the open-string tachyon field $\tau$ models chiral symmetry breaking holographically. $\hat{F}$ is the field strength for the baryon number gauge field. In this work, we are interested in a chirally symmetric phase of the quark-gluon plasma at zero density, for which the tachyon and gauge field are both set to zero $\tau = \hat{F} = 0$. We may work in this case with a reduced flavor action
\begin{equation}\label{flavacred}
S_{flavor}=-x_fM_p^3N_c^2\int d^5x V_{f0}\lr{\Phi}\sqrt{-\text{det}g}\, .
\end{equation}

The flavor action \eqref{flavacred} depends on an additional dilaton potential $V_{f0}\lr{\Phi}$. Small and large $\Phi$ asymptotics of this potential follow from the requirement to reproduce a number of salient features of QCD in the mesonic sector. Given these constraints, there remains some indeterminacy in the precise form of the potential, which is however limited by fitting both heavy-ion collision experiments and neutron star mergers, see \cite{Jarvinen:2021jbd,Demircik:2021zll,Jokela:2021vwy,Hoyos:2021njg,Ecker:2019xrw,Jokela:2018ers,Alho:2015zua,Alho:2013hsa,Arean:2012mq}. Based on these fits we make the following choice for the glue and flavor potentials: 
\begin{equation}\label{Vgf}
\begin{split}
    &V_g\lr{\Phi}=12\left[1+V_1\lambda+\frac{V_2\lambda^2}{1+\lambda/\lambda_0}+V_{IR}\,e^{-\lambda_0/\lambda}\lr{\lambda/\lambda_0}^{4/3}\sqrt{\text{log}\lr{1+\lambda/\lambda_0}}\right],\\
    &V_{f0}\lr{\Phi}=W_0+W_1\lambda+\frac{W_2\lambda^2}{1+\lambda/\lambda_0}+12 W_{IR}e^{-\lambda_0/\lambda}\lr{\lambda/\lambda_0}^{2}\, ,
\end{split}
\end{equation}
where we defined $\lambda = \exp{\Phi}$ to simplify notation. 
It is important to stress that all the parameters that appear above are fixed. On the UV side, they are matched to the RG flow of QCD perturbation theory, and on the IR side they are determined by comparing to lattice data. Below we present the parameter values that we use, corresponding to the potential set 7a in \cite{Jarvinen:2021jbd} which is a particular V-QCD model with intermediate stiffness of the equation of state:
\begin{equation}\label{params}
\begin{split}
   & V_1=\frac{11}{27\pi^2}\, , \quad V_2=\frac{4619}{46656\pi^4}\, , \quad W_1=\frac{8+3W_0}{9\pi^2}\, , \quad W_2=\frac{6488+999W_0}{15552\pi^4} \\ \\ 
   &\lambda_0=8\pi^2/3\, , \quad V_{IR}=2.05\, , \quad W_0=2.5 \, , \quad W_{IR}=0.9 \, , \quad \Lambda_{UV}=211\text{MeV}\,,
\end{split}
\end{equation}
with $\Lambda_{UV}$ setting the scale of the boundary theory. The effective potential for V-QCD that enters the semi-analytic formulas for shear and bulk viscosity per entropy, given in section \ref{sec::gen}, is the combination 
\begin{equation}\label{Veff}
    V_{eff}\lr{\Phi}=V_g \lr{\Phi}-x_f  V_{f0}\lr{\Phi}\, .
\end{equation}
In the following, the ratio $x_f=N_f/N_c$ is set to 1. Figure \ref{VQCD dilaton potential and coupling} provides a plot of the effective V-QCD potential \eqref{Veff} in this case, with potential parameters chosen as in \eqref{params}.  

Our setup is different from usual V-QCD, since it includes a higher derivative term in the glue action (\ref{gulac}). In the following, it is assumed that this term only leads to small order $\mathcal{O}(\beta)$ corrections to the background, which do not alter dramatically the properties of the holographic-QCD theories. In particular, we require the IR asymptotics to preserve the confinement criteria established in \cite{Gursoy:2007er}. The ${\cal O}(\beta)$ corrections to both the UV and IR asymptotics are worked out in appendix \ref{sec::bg}, where it is shown in particular that confinement is preserved for a generic ansatz 
\begin{equation}\label{GIR}
G\lr{\Phi}=G_{IR}e^{\gamma\Phi}\Phi^{\delta}\,,\qquad \gamma\leq -\frac43\, .
\end{equation} 

\section{Fitting holographic models to Bayesian analyses}
\label{sec::fit}

\subsection{Shear viscosity}

\label{ses}

We explained at the end of section \ref{sec::gen} how to fit any desired temperature profile for $\eta/s$ by appropriately choosing the function $G(\Phi)$. However, the profiles that follow from the Bayesian analysis, see Figure \ref{exp visc figure}, contain wide errors allowing for a wide range of choices for this function. We therefore start with the simplest choice, $G=1$, and ask whether the data can be fitted within the 90\% confidence bands. 

\begin{figure}[h]
    \centering
    \includegraphics[width=0.65\textwidth]{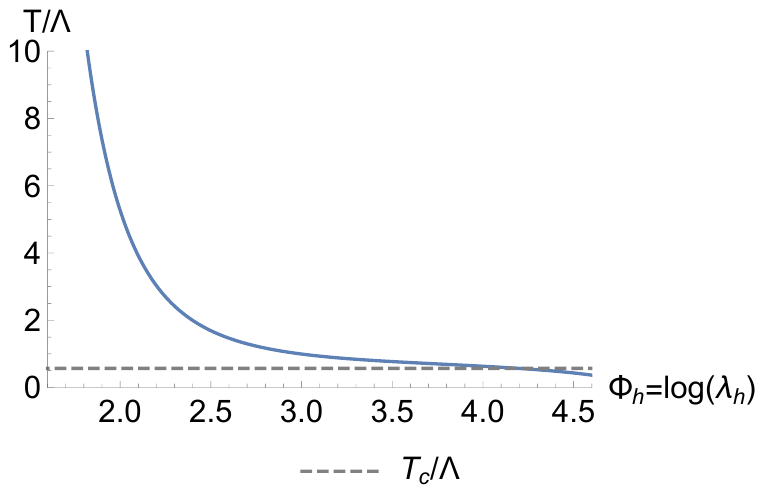}
    \caption{Temperature in units of the confining scale $\Lambda$, as a function of the dilaton horizon value in V-QCD. The dotted line indicates the value of the critical temperature $T_c/\Lambda$, where the deconfining transition happens.}
    \label{T(phi_h)}
\end{figure}
 
\begin{figure}[h]
  \centering
  \begin{minipage}[b]{0.45\textwidth}
    \includegraphics[width=\textwidth]{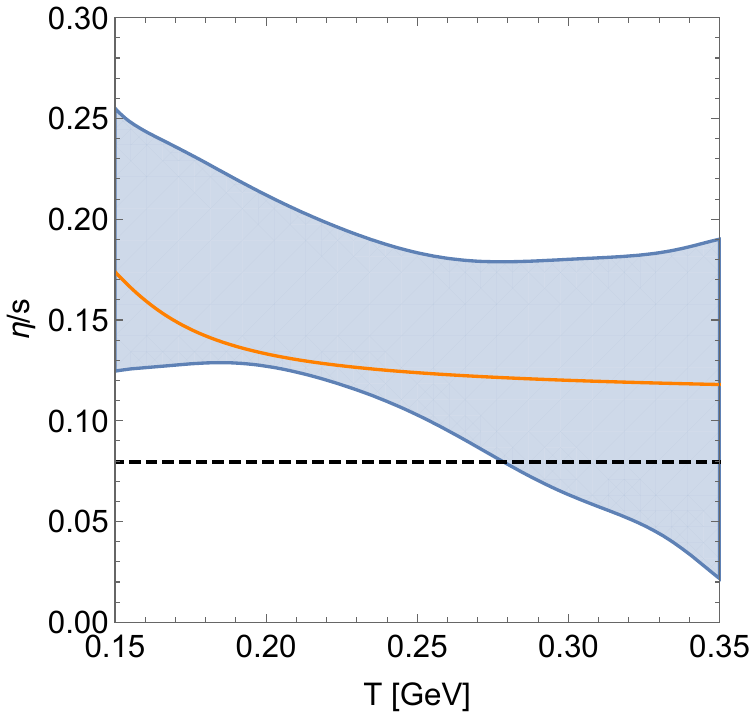}
  \end{minipage}
  \hfill
  \begin{minipage}[b]{0.45\textwidth}
    \includegraphics[width=\textwidth]{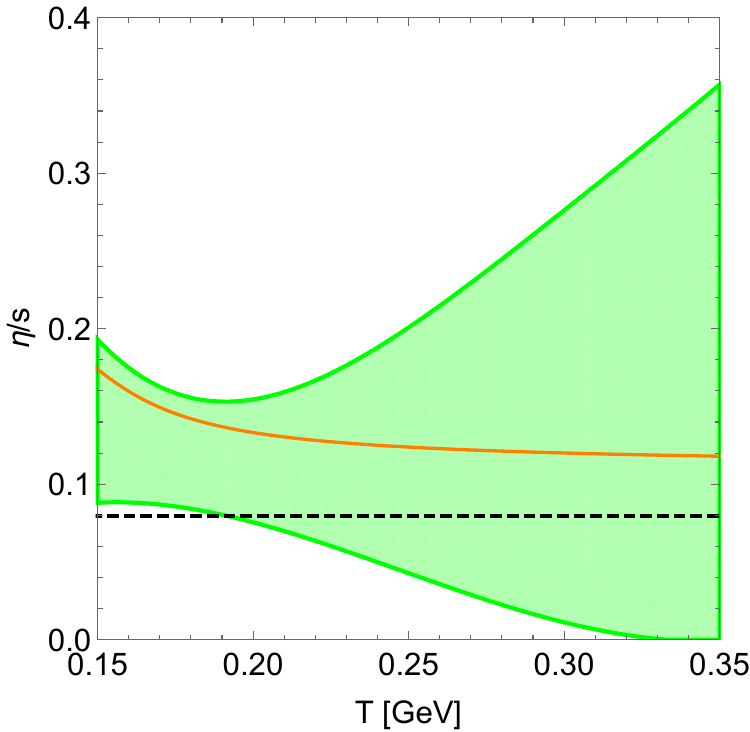}
  \end{minipage}
   \caption{Shear viscosity to entropy ratio for V-QCD with  $G\lr{\Phi}=1$. The blue area corresponds to the 90$\%$ confidence band for Trajectum while the green area corresponds to the 90$\%$ confidence band for Jetscape. The orange line shows the theoretical curve while the black line is $\eta/s=1/(4\pi)$ for reference. A curve that fits both Trajectum and Jetscape is obtained for $\beta=-0.05$.}
   \label{VQCD model1}
\end{figure}

For $G=1$, (\ref{eta over s theory}) reduces to 
\begin{equation}
    \frac{\eta}{s}=\frac{1}{4\pi}\left[1-\frac{2}{3}\beta V\lr{\Phi_h}\right]\, ,
\end{equation}
were we set $\ell=1$. Since we work with a fixed V-QCD potential $V_{eff}\lr{\Phi_h}$ (see \eqref{Vgf}-\eqref{Veff}), the only free parameter is $\beta$. Assuming $\beta \ll 1$, the relation between the temperature and $\Phi_h$ can be determined from the black-brane surface gravity omitting the curvature corrections. The resulting profile for $T(\Phi_h)$ is shown in Figure \ref{T(phi_h)}, while Figure \ref{VQCD model1} shows the $\eta/s$ temperature profile for the value of $\beta$ that gives the best fit to data. This indicates that a constant $G\lr{\Phi}$ is compatible with both sets of Bayesian data. Even though the resulting curve is not very close to the central values inferred from Bayesian analyses, it definitely lies within the 90\% confidence range. It will be interesting to observe whether this continues to hold when future data allow to compute narrower confidence bands.  

We now move on to the case of general $G(\Phi)$ and solve the differential equation (\ref{eta over s theory}) numerically to determine the optimal fit to Bayesian data. As an example we consider the data from Trajectum, see Figure \ref{exp visc figure}. 
\begin{figure}[h]
    \centering
    \includegraphics[width=0.5\textwidth]{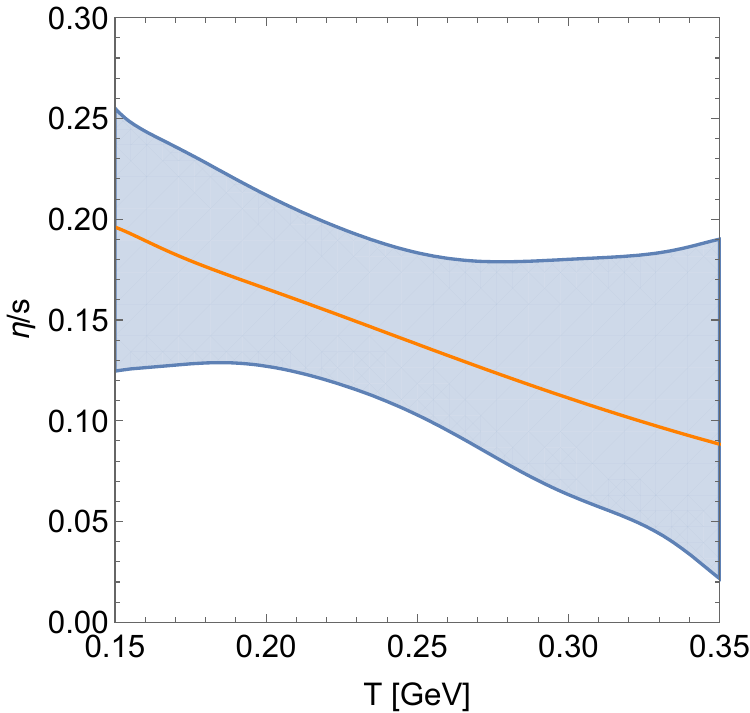}
    \caption{Best fit for the shear viscosity to entropy ratio for V-QCD with  $G\lr{\Phi}$ shown in Figure \ref{VQCD dilaton potential and coupling} ($\beta$ is set to 0.1). The orange line is the theoretical curve, which is compared with the 90$\%$ confidence band for Trajectum (blue area).}
    \label{VQCD model2}
\end{figure}
We set $\beta=0.1$ without loss of generality and compute the function $G$ which reproduces the mean of the Bayesian data, shown as the orange curve in Figure \ref{VQCD model2}. Equation \eqref{G solution} gives the corresponding potential $G(\Phi)$, which is plotted in Figure \ref{VQCD dilaton potential and coupling} (right), along with the effective V-QCD potential $V_{eff}(\Phi)$ (left). 
\begin{figure}[h]
  \centering
  \begin{minipage}[b]{0.495\textwidth}
    \includegraphics[width=\textwidth]{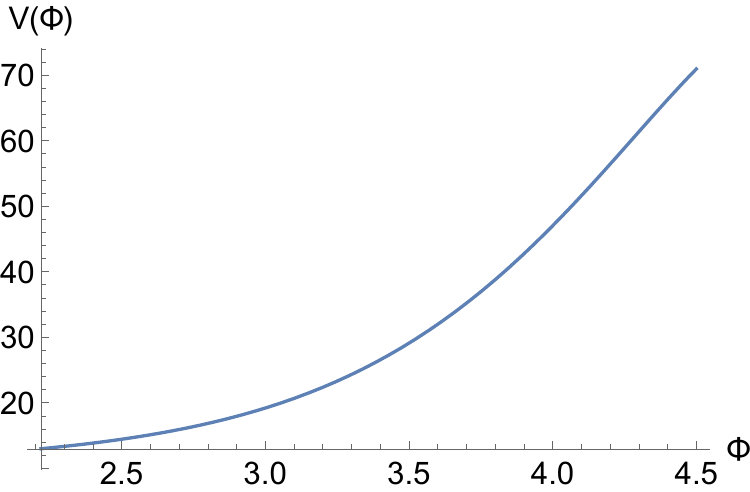}
  \end{minipage}
  \hfill
  \begin{minipage}[b]{0.495\textwidth}
    \includegraphics[width=\textwidth]{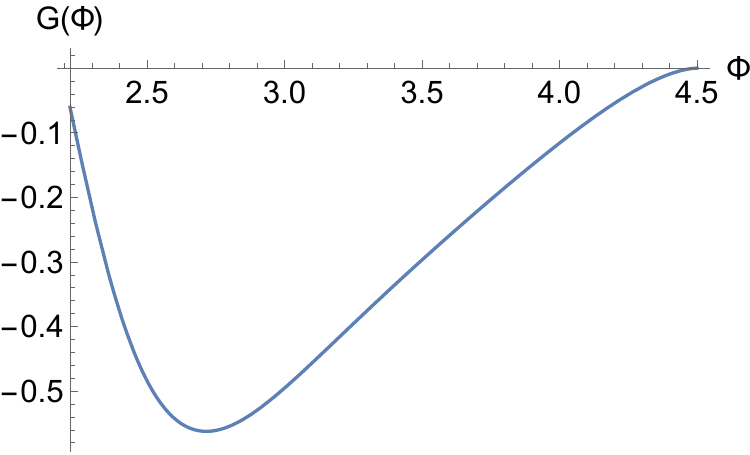}
  \end{minipage}
   \caption{Dilaton potentials for the model with $\eta/s$ as in Figure \ref{VQCD model2}.}
   \label{VQCD dilaton potential and coupling} 
\end{figure}
A few remarks are in order: 
\begin{itemize} 
\item As mentioned above, the integration constant $c_1$ in \eqref{G solution} is irrelevant to $\eta/s$. It will matter however for the fit to bulk viscosity. In Figure \ref{VQCD dilaton potential and coupling}, $c_1$ was chosen\footnote{This choice was made such that $G$ is flat for large $\Phi$ hence the derivatives remain small, so that we stay within the perturbative regime.} such that $G\lr{\Phi_h=4.5}=0$.
\item Outside of the fitting region shown in Figure \ref{VQCD dilaton potential and coupling}, we make sure that G drops off to zero according to the asymptotic behaviors derived in appendix \ref{sec::bg}. These behaviors ensure that the order $\mathcal{O}(\beta^0)$ asymptotics of the background and the associated QCD-like properties of the model are not modified by the curvature corrections.
\item More importantly, we find that the correction to the background solution for this choice of $G$ violates our perturbative treatment. This means that, to fit Bayesian data with our holographic model, it is necessary to solve the metric equations including the curvature corrections. This indicates that finite coupling corrections $1/\lambda_{QCD}$ are significant. Below, we present an alternative fitting scheme for both shear and bulk viscosities which agrees less well with the Bayesian data but remains within the perturbative regime. 
\end{itemize}
\subsection{Bulk viscosity}

In this section, we focus on comparison of the bulk viscosity per entropy with the Bayesian analysis. For simplicity, we again consider the V-QCD action with a single Riemann-square type correction with coupling $\beta$ given by $S_{glue} + S_{flavor}$ (see (\ref{gulac}) and (\ref{flavac})).   

\subsubsection{Bulk viscosity of V-QCD theory}

As a first step we calculate $\zeta/s$ for the original V-QCD theory with no curvature corrections, i.e. with $\beta=0$. The easiest way to compute the bulk viscosity in this case is to employ the Eling-Oz formula \cite{Eling:2011ms}:  
 \begin{equation}
    \frac{\zeta}{s}=\frac{2}{3\pi}\lr{s\frac{\partial\Phi_h}{\partial s}}^2 \,. 
\end{equation}
Equivalence between this formula and the standard holographic computation through fluctuating the background has been established in \cite{Demircik:2023lsn}. 

\begin{figure}[h]
\centering
\includegraphics[width=0.55\textwidth]{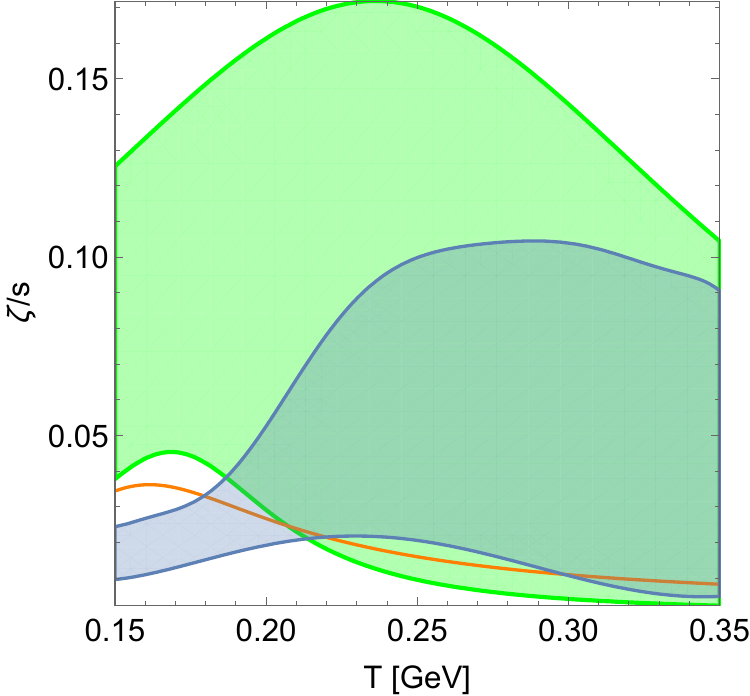}
    \caption{Bulk viscosity to entropy ratio as a function of temperature for V-QCD without curvature corrections. The orange line is the model prediction. The green region is the 90\% confidence band from the Jetscape collaboration and is derived from combined RHIC and LHC data. The blue region comes from the analysis of the Trajectum collaboration.}
    \label{bulk visc temp data}
\end{figure}

The resulting V-QCD bulk viscosity is plotted in Figure \ref{bulk visc temp data} as a function of temperature, along with the 90$\%$ confidence bands of the two data sets.
Some comments about this fit:
\begin{itemize}
\setlength\itemsep{1em}
    \item The fit has no additional free parameters since the V-QCD potential used is completely fixed by other requirements; see e.g. \cite{Jarvinen:2021jbd}.
    \item This zeroth-order result seems to agree better with the data from Jetscape. In particular, the location of the maximum at $T\simeq 0.16\,\mathrm{GeV}$ does not fit well with the Trajectum results.
\end{itemize}

\subsubsection{Bulk viscosity with curvature corrections}

\begin{figure*}[t!]
        \subfloat{%
    \hspace{-0.2cm}{\includegraphics[width=.495\linewidth]{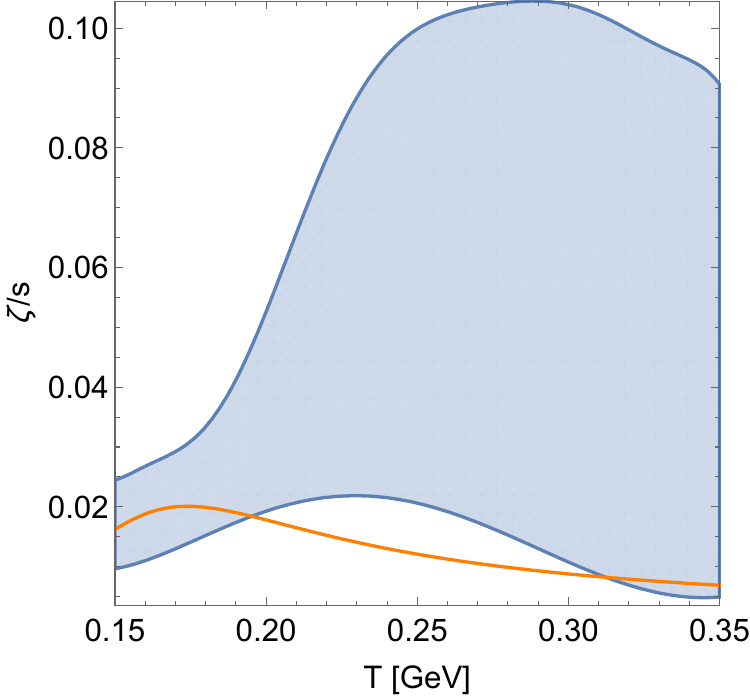}}
            \label{}%
        }
        \subfloat{%
        \includegraphics[width=.495\linewidth]{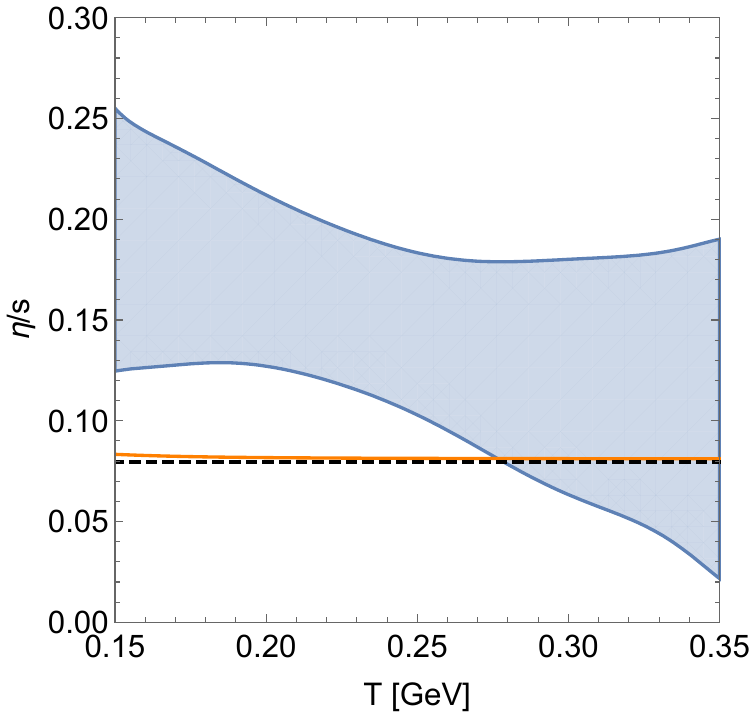}%
            \label{}%
        }\\
        \subfloat{%
        \includegraphics[width=.495\linewidth]{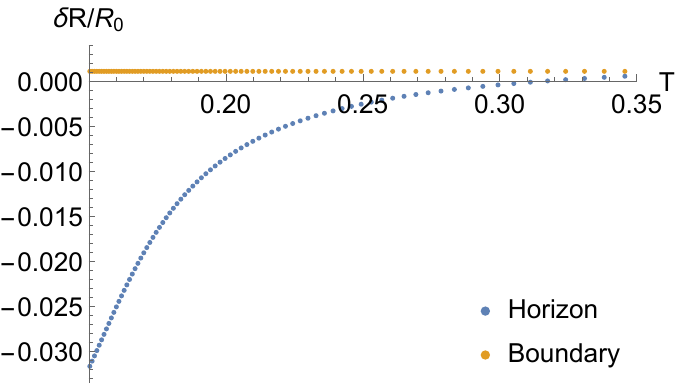}%
            \label{}%
        }\hfill
        \subfloat{
        
        \includegraphics[width=.495\linewidth]{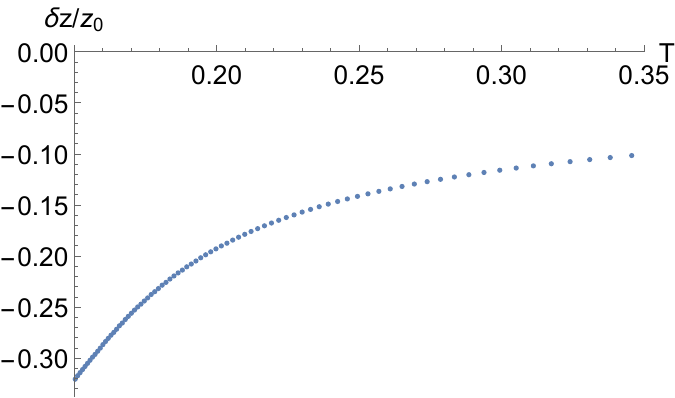}
            \label{}%
        }
         \caption{Bulk and shear viscosities for $\beta G=-0.002$ compared with Trajectum and explicit check of the magnitude of the corrections for the background and bulk perturbations. The bottom left plot is the relative change in the Ricci scalar near the horizon and boundary. The bottom right plot is the relative change in the gauge invariant dilaton fluctuations $Z$ at the horizon. $Z$ is defined in \eqref{defZ} and its dynamics determines the bulk viscosity. Both relative differences are largest in absolute value at the horizon. In all figures the temperature is in GeV.}
   \label{bulk trajectum fit}
\end{figure*}

We now discuss the effect of curvature corrections on the V-QCD bulk viscosity per entropy. In particular, we would like to investigate whether these corrections can bring $\zeta/s$ closer to either of the bands from the Bayesian analyses. As for the shear viscosity, we can obtain some insight by starting from the simplest form of curvature corrections, namely a Riemann-squared term with constant $G$. The Eling-Oz formula has not yet been generalized to include higher curvature corrections; therefore, we need to follow the standard holographic method for this calculation, based on fluctuation of the background \cite{Buchel:2023fst}. Details can be found in appendix \ref{sec::appD} and \ref{sec::appE}.

\begin{figure*}[t!]
        \subfloat{%
        \hspace{-0.2cm}\includegraphics[width=.49\linewidth]{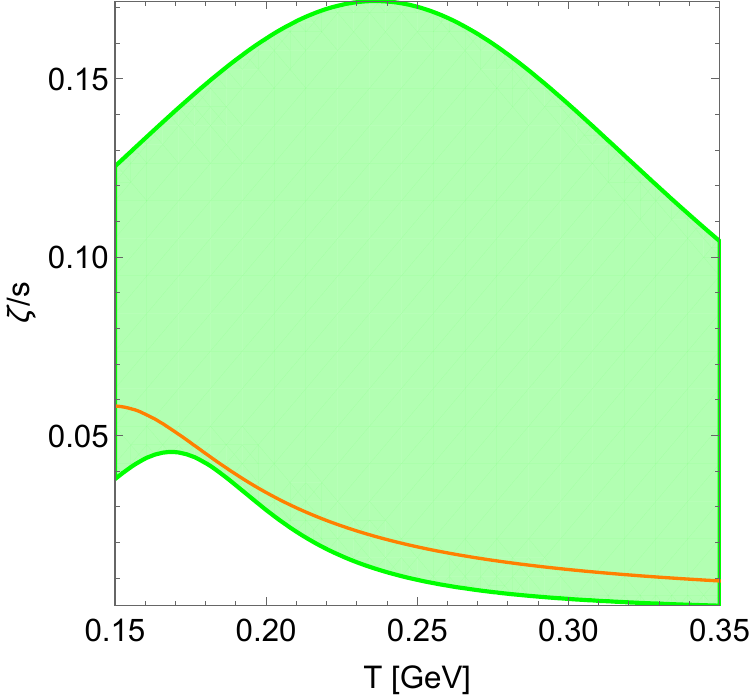}%
            \label{}%
        }\hfill
        \subfloat{%
        \includegraphics[width=.48\linewidth]{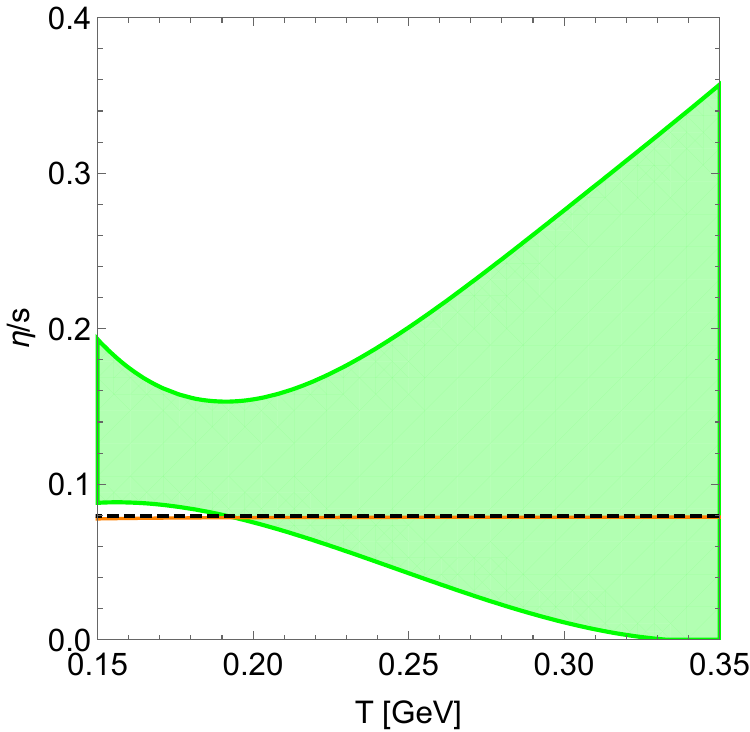}%
            \label{}%
        }\\
        \subfloat{%
        \includegraphics[width=.49\linewidth]{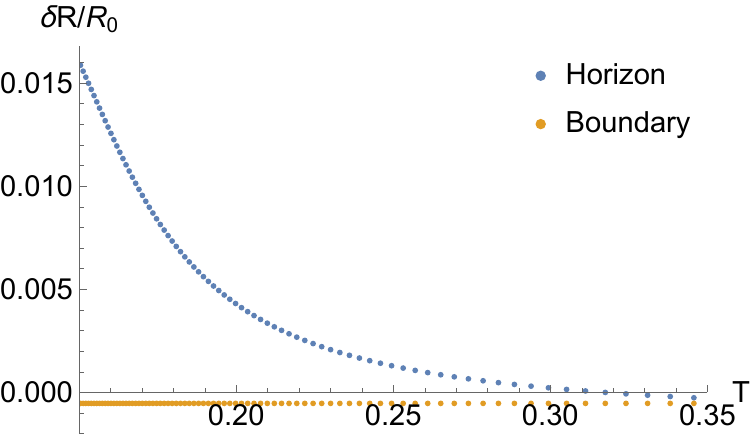}%
            \label{}%
        }\hfill
        \subfloat{%
        \includegraphics[width=.48\linewidth]{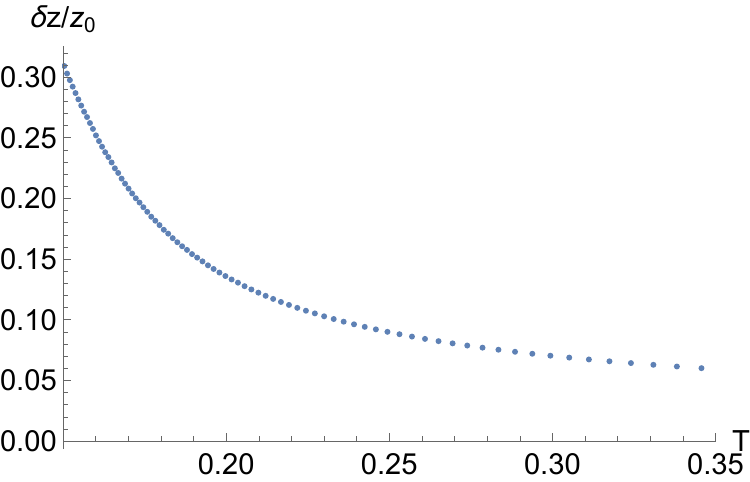}%
            \label{}%
        }
         \caption{Same as Figure \ref{bulk trajectum fit}, but for $\beta G=0.001$ and comparing with Jetscape results.}
   \label{bulk jetscape fit}
\end{figure*}

Figure \ref{bulk trajectum fit} shows the temperature profile of $\zeta/s$ for $\beta G = -0.002$. The latter is seen to shift closer to Trajectum at low temperatures while remaining within the regime of applicability of our perturbative calculation, as shown in the bottom figures. However, the profile remains outside of the Trajectum band at intermediate temperatures $0.2\,\mathrm{Gev}\lesssim T \lesssim 0.3\,\mathrm{GeV}$. 

On the other hand, the temperature profile of $\zeta/s$ can be made compatible with the Jetscape band if we consider positive values of $\beta G$. This is shown in Figure \ref{bulk jetscape fit}, where we set $\beta G = 0.001$. The bottom plots indicate that this calculation also consistently remains within the perturbative regime.

As an indication, Figures \ref{bulk trajectum fit} and \ref{bulk jetscape fit} also show the corresponding profiles of $(\eta/s)(T)$, which are nearly unchanged from $1/(4\pi)$ by the small derivative corrections.

\subsection{Global fits to both shear and bulk viscosity data}

The analysis of the previous subsection shows that the viscosities can be brought close to Jetscape data with a constant coefficient $\beta G$ for the curvature corrections. However, we also observed that the tension with Trajectum data which exists at zeroth-order cannot be resolved by this kind of simple curvature corrections. In this section, we perform a joint analysis of the shear and bulk viscosity, with the motivation to find potentials $G(\Phi)$ more appropriate for fitting Trajectum results.       

\begin{figure*}[t!]
        \subfloat{%
        \hspace{0.3cm}\includegraphics[width=.49\linewidth]{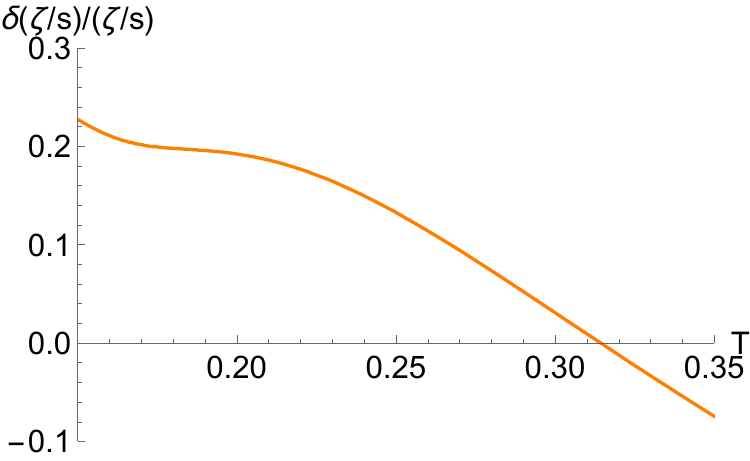}%
            \label{}%
        }\hfill
        \subfloat{%
        \includegraphics[width=.48\linewidth]{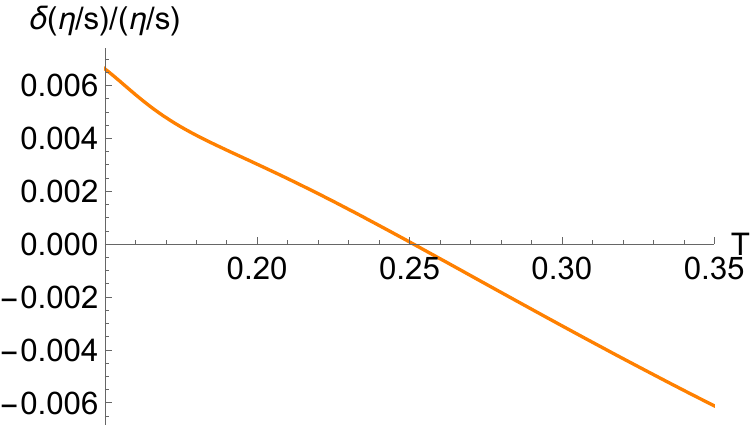}%
            \label{}%
        }\\
        \subfloat{%
        \includegraphics[width=.49\linewidth]{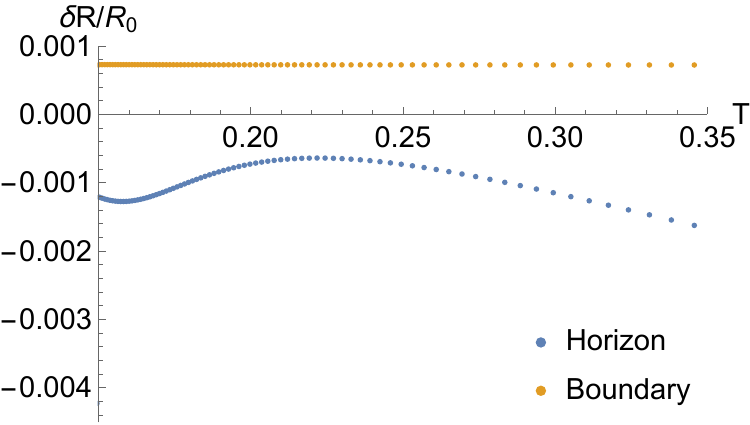}%
            \label{}%
        }\hfill
        \subfloat{%
        \includegraphics[width=.48\linewidth]{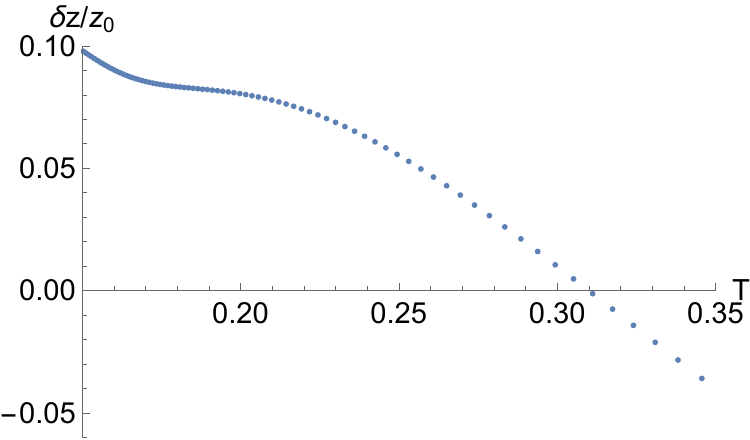}%
            \label{}%
        }
         \caption{Relative change in the bulk and shear viscosities with the $\mathcal{O(\beta)}$ result of V-QCD for a linearly decreasing $\eta/s$ \eqref{etalin} with a slope $\gamma = -0.005$. The bottom plots are the same type of relative differences as in figure \ref{bulk trajectum fit}. The temperature is in GeV.}
   \label{linear viscosity fit}
\end{figure*}

The Trajectum analysis suggests that $\eta/s$ decreases with temperature in the range of interest. Based on this observation, we will assume a linear ansatz for $(\eta/s)(T)$ 
\begin{equation}\label{etalin}
\eta/s=1/4\pi+\gamma(T-T^{*}) \, , 
\end{equation}
with $T^*=0.25\,\mathrm{GeV}$ in the middle of the temperature range of interest, and $\gamma$ parametrizing the slope of the temperature profile, which should be small enough for our perturbative treatment to apply. Specifically, we find that a choice $|\gamma|\lesssim 0.005$ allows to remain  within the perturbative regime. The potential $G(\Phi)$ that generates \eqref{etalin} is then fixed from \eqref{eta over s theory}, and the goal is to observe the qualitative change in bulk viscosity with this kind of potential, for which $\eta/s$ decreases with temperature. With $|\gamma|\lesssim 0.005$, the shear viscosity profile is of course close to a constant, and very different from the Trajectum central value. However, computing the corresponding profile for $\zeta/s$ gives an indication whether it can get closer to Trajectum data in this more realistic scenario; in particular whether the maximum of $\zeta/s$ tends to move towards higher temperatures.   

The order $\mathcal{O}(\beta)$ relative differences in viscosities resulting from this calculation are shown in Figure \ref{linear viscosity fit}, together with the corrections to the background. Unlike the constant $G$ case, the qualitative change in $\zeta/s$ is observed to depend on temperature: it increases at low temperatures but decreases at higher temperatures. There is therefore no indication of $\zeta/s$ approaching the Trajectum data when $\eta/s$ is imposed to decrease linearly with temperature, although a fully non-perturbative analysis may lead to different conclusions. This suggests that fitting both types of viscosities requires including other types of curvature corrections ($R^2$ and $R_{\m\n}^2$ at quadratic order), at least for the Trajectum case.  
    
As a final note, we would like to discuss the alternative of working with different holographic models. In particular, demanding that the effect of higher-curvature corrections on the background is negligible, the model used here seems to fit the Jetscape data better (which is alsmost compatible with $\eta/s=1/(4\pi)$). It is then natural to ask whether one can fit the two separate data sets using different holographic models. In particular, it is tempting to use a constant potential $V$ for Trajectum, such that the theory is conformal at order $\mathcal{O}(\beta^0)$ with a vanishing bulk viscosity. Since Trajectum data is consistent with $\zeta/s \lesssim 0.02$, such a conformal model may give a reasonable approximation for this case.  

\begin{figure*}[t!]
\centering
        \subfloat{%
        \includegraphics[width=.49\linewidth]{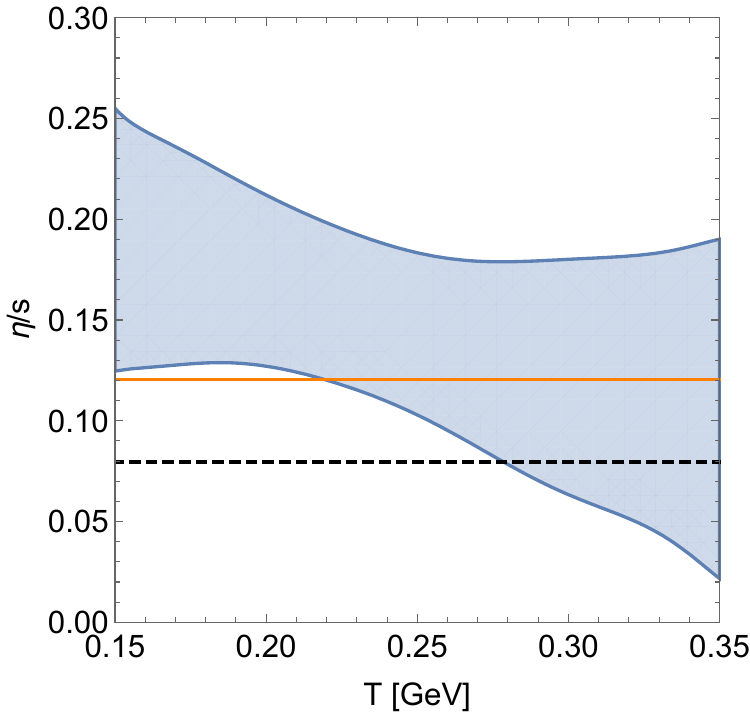}%
            \label{}%
        }\hfill
        \subfloat{%
        \raisebox{1.2cm}{\includegraphics[width=.49\linewidth]{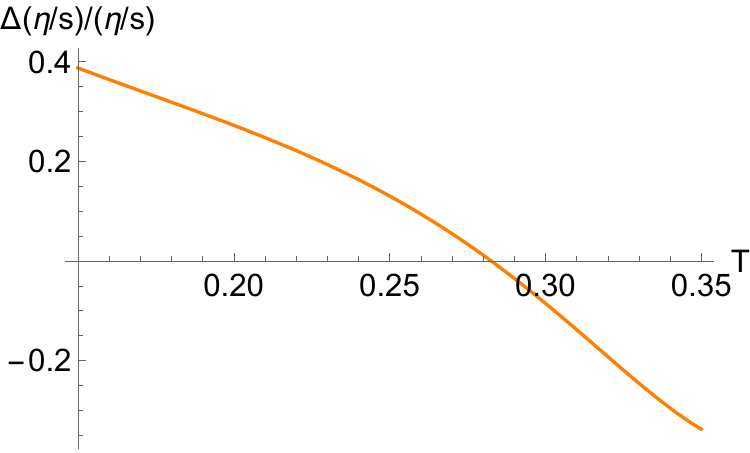}}
            \label{}
        }
         \caption{Shear viscosity to entropy ratio for an AdS solution with constant higher derivative corrections. The value of $\beta =-0.064$ was picked such that it provides the best fit to the Trajectum central value of $\eta/s$, while at the same time requiring that the change in the background Ricci scalar induced by the higher curvature terms should obey $\delta R/R<0.1$. The right plot is the relative difference in $\eta/s$ with the central Trajectum value, $[(\eta/s)_{\text{Traj}}-\eta/s]/(\eta/s)_{\text{Traj}}$, as a function of temperature in GeV.}
   \label{eta/s AdS fits}
\end{figure*}

Taking $G$ to be a constant preserves conformal invariance at the higher derivative level, so that we only need to fit $\eta/s$. Constant $V$ and $G$ imply a constant $\eta/s$, which is compatible with Trajectum data, but quite different from the linearly decreasing central value. This is shown in more detail in Figure \ref{eta/s AdS fits}, which compares the Trajectum central value $(\eta/s)_{\text{Traj}}$ with the best constant fit compatible with a perturbative treatment of the curvature corrections\footnote{Our criterium is such that the change in the background Ricci scalar induced by the curvature squared terms should obey $\delta R/R < 10\%$.}. In particular, the relative difference $[(\eta/s)_{\text{Traj}}-\eta/s]/(\eta/s)_{\text{Traj}}$ (right Figure \ref{eta/s AdS fits}) is seen to exceed $30\%$ at the edges of the fitting range.

\section{Discussion and Outlook}
\label{sec:dis}
We explored various possibilities to fix the curvature corrections in the realistic holographic QCD model, V-QCD, in a way consistent with the temperature dependence of shear and bulk viscosity observed in the Bayesian analysis of heavy-ion collisions \cite{Nijs:2020ors,Nijs:2020roc,Nijs:2023yab,Giacalone:2023cet,JETSCAPE:2020mzn}. 

The curvature corrections were treated in a perturbative manner, which allowed us to work with second order equations of motion, instead of the full fourth order problem. Solving the fourth order equations is in principle possible. However, it is well known that higher derivative terms in metric equations may lead to causality violation, see e.g. \cite{Buchel:2009tt,Brigante:2008gz}. It would be interesting to explore the causality of the 4-derivative corrections to holographic-QCD theories in general. Staying within the causal range, it would make it possible to go beyond the perturbative treatment. This could provide more flexibility in fitting both viscosity profiles. 

However, one should then also ask whether such corrections are compatible with the previously established properties of ihQCD or V-QCD holographic models i.e., confinement, desired running of the 't Hooft coupling, asymptotics of glueball and meson spectra, chiral symmetry breaking, holographic realization of QCD anomalies, properties of the phase diagram at finite temperature and chemical potential, etc. In principle, we would need to make another global fit including both the potentials related to curvature corrections, $G_1$, $G_2$ and $G_3$ and also the other potentials of V-QCD i.e. $V_g$, $V_{f0}$, $\kappa$ and $w$, with the requirement of the aforementioned properties. 

Even if we manage to reproduce the desired QCD features, there remains the problem of other higher curvature corrections in a non-perturbative treatment. That is, once we treat the four-derivative corrections as the same order as the original background, then why not include higher derivative corrections. One possible answer is to treat the theory with four-derivative corrections only, as a phenomenological model, which fits well some QCD observables (for example the viscosities per entropy), but is probably less good at reproducing other QCD properties away from the IR regime. Another possibility is to try to include a whole series of higher derivative corrections to the 5D non-critical string theory altogether by introducing an $O(d,d)$ invariant action as in the double field theory; see e.g. \cite{Hohm:2010jy,Hohm:2010pp}. In this case, one could also employ the constraints that arise in the ``swampland'' program \cite{Vafa:2005ui,Ooguri:2006in}, see e.g. \cite{Palti:2019pca} for a review, which in general restricts the range of consistent UV-complete higher derivative theories. All these questions are clearly interesting but beyond the scope of this work. 

Apart from these structural questions, an interesting extension of our work involves charge transport. Presence of dense magnetic fields in off-central collisions results in various charge-transport related effects in the quark-gluon plasma with observable consequences; see for example  \cite{Kharzeev:2007jp,Skokov:2009qp,Tuchin:2013ie,Voronyuk:2011jd,Deng:2012pc,Tuchin:2013apa,McLerran:2013hla,Gursoy:2014aka,Gursoy:2018yai}, and  \cite{Kharzeev:2013jha} for a review. Indeed, extensions of the Bayesian analyses to include charge transport coefficients and the presence of electromagnetic fields in off-central events would, most importantly, also provide a temperature profile for conductivity, $\sigma$. Now, there is also a universal formula for conductivity which leads to a model-independent constant for the ratio of $\sigma$ and $s^{1/3}$ in two-derivative Einstein gravity \cite{Iqbal:2008by}. To go beyond this and model a more generic temperature dependence in holography, one also needs higher derivative corrections to holographic setups which include magnetic fields e.g. \cite{DHoker:2009mmn,Ballon-Bayona:2013cta,Rougemont:2015oea,Finazzo:2016mhm,Gursoy:2017wzz,Gursoy:2020kjd}; see \cite{Gursoy:2021efc} for a review. We wish to turn back to the points listed here in the future.

\section*{Acknowledgements}
We would like to thank Alex Buchel and Sera Cremonini for useful correspondence. We are also particularly grateful to Wilke van der Schee and Govert Nijs for sharing their data generated with Trajectum. EP and UG are supported by the Netherlands Organization for Scientific Research (NWO) under the VICI grant VI.C.202.104. Work of UG is also supported by the ENW-XL program ``Probing the QCD phase diagram" of NWO.

\clearpage
\appendix

\section{The vacuum solution}
\label{sec::bg}

We discuss in this appendix the zero-temperature solution corresponding to the vacuum of the theory. We start from presenting the equations of motion that it solves, before discussing the behavior in the UV and IR asymptotic limits. As mentioned in the main text, the asymptotics of the V-QCD potentials are chosen to reproduce several salient qualitative features of QCD. We describe how requiring that those features are not spoiled by the higher curvature corrections puts constraints on the additional potential $G(\Phi)$.  In the following, we work with ihQCD but the extension to V-QCD is straightforward. 

\subsection{Equations of motion}

The equations of motion are derived from the action \eqref{gulac}. The dilaton equation takes the form
\begin{equation}\label{EOMF}
    \frac{8}{3}\nabla_{\mu}\nabla^{\mu}\Phi+\ell^2\beta R_{\mu\nu\rho\sigma}R^{\mu\nu\rho\sigma}\partial_{\Phi}G\lr{\Phi}+\partial_{\Phi}V\lr{\Phi}=0\, ,
\end{equation}
whereas the metric equation is given by 
\begin{equation}\label{EOMg}
     \begin{aligned} & R_{\mu\nu}-\frac{1}{2} g_{\mu\nu}R-\frac{1}{2} g_{\mu\nu}V\lr{\Phi}-\frac{4}{3}\left(\nabla_{\mu} \Phi\right)\left(\nabla_{\nu} \Phi\right)+\frac{2}{3} g_{\mu\nu}\left(\nabla_{\sigma} \Phi\right)\left(\nabla^{\sigma} \Phi\right)+ \\ & 
     +\ell^2 \beta\bigg[2 G\lr{\Phi} R_{\mu}^{\;\;\rho \sigma \alpha} R_{\nu \rho \sigma \alpha}-\frac{1}{2} g_{\mu \nu} G\lr{\Phi} R_{\mu \nu \rho \sigma} R^{\mu \nu \rho \sigma}+2 G\lr{\Phi}\nabla^{\rho} \nabla^{\sigma} R_{\mu\rho \nu\sigma}+ \\ 
     & +2 G\lr{\Phi}\nabla^{\sigma} \nabla^{\rho} R_{\mu\rho \nu\sigma}+4\left(\nabla^{\rho} \Phi\right)\nabla^{\sigma} R_{\mu \rho \nu \sigma}\partial_{\Phi}G\lr{\Phi}+4\left(\nabla^{\rho} \Phi\right)\left(\nabla^{\sigma} R_{\mu \sigma \nu \rho}\right) \partial_{\Phi}G\lr{\Phi}+ \\ 
     & +2 R_{\mu \rho \nu \sigma}\left(\nabla^{\sigma} \nabla^{\rho} \Phi\right) \partial_{\Phi}G\lr{\Phi}+2 R_{\mu \rho \nu \sigma}\left(\nabla^{\rho} \nabla^{\sigma} \Phi\right) \partial_{\Phi}G\lr{\Phi}+\\
     &+4 R_{\mu \rho \nu \sigma}\left(\nabla^{\rho} \Phi\right)\left(\nabla^{\sigma} \Phi\right) \partial_{\Phi}^{2}G\lr{\Phi}\bigg]=0 , \end{aligned}
\end{equation}
where $V(\Phi)$ is the ihQCD dilaton potential. 

The background solutions that we are interested in are homogeneous, horizon-less configurations, for which an appropriate ansatz is given by
\begin{equation}
\label{T=0 metric}
    ds^2=e^{2A\lr{r}}\lr{dr^2-dt^2+d\mathbf{x}^2} \, , \quad  \Phi=\Phi\lr{r} \, ,
\end{equation}
where $r$ is the holographic coordinate, which goes to zero at the boundary. Substituting this ansatz into the equations of motion \eqref{EOMF}-\eqref{EOMg} results in the following set of equations for the ansatz fields $A(r)$ and $\Phi(r)$
\begin{equation}
\label{rr and tt components}
    \begin{aligned} & 108 \ell^2 \beta G\lr{\Phi} A'\lr{r}^4-36 \ell^2 \beta A'\lr{r}^3 \partial_{\Phi}G\lr{\Phi} \Phi'\lr{r}-9 e^{2 A\lr{r}} A''\lr{r}+\\
    &+48 \ell^2 \beta G\lr{\Phi} A''\lr{r}^2+ 3A'\lr{r}^2\left(3 e^{2 A\lr{r}}-52 \ell^2 \beta G\lr{\Phi} A''\lr{r}\right)-\\&
    -4 \Phi'\lr{r}^2\left(e^{2 A\lr{r}}-3 \ell^2 \beta A''\lr{r} \partial_{\Phi}^{2}G\lr{\Phi}\right) \\ &+12 \ell^2 \beta \partial_{\Phi}G\lr{\Phi} A''\lr{r} \Phi''\lr{r} +24 \ell^2 \beta \partial_{\Phi}G\lr{\Phi} \Phi'\lr{r} A^{(3)}\lr{r}-\\
    &-36 \ell^2 \beta A'\lr{r}\big(\partial_{\Phi}G\lr{\Phi} \Phi'\lr{r}A''\lr{r} +G\lr{\Phi}A^{(3)}\lr{r}\big)+ 12 \ell^2 \beta G\lr{\Phi} A^{(4)}\lr{r}=0 \, , \end{aligned}
\end{equation}
\begin{equation}
\label{rr - tt components}
\begin{aligned}
    &e^{4 A\lr{r}} V\lr{\Phi}-3 e^{2 A\lr{r}} A''\lr{r}-A'\lr{r}^2\left(9 e^{2 A\lr{r}}+20 \ell^2 \beta G\lr{\Phi} A''\lr{r}\right)-\\ 
    &-12 \ell^2 \beta A'\lr{r}^3 \partial_{\Phi}G\lr{\Phi} \Phi'\lr{r}+4 \ell^2 \beta \Phi'\lr{r}^2A'\lr{r} \partial_{\Phi}^{2}G\lr{\Phi}+4 \ell^2 \beta \partial_{\Phi}G\lr{\Phi} A''\lr{r} \Phi''\lr{r}+ \\
    &+8 \ell^2 \beta \partial_{\Phi}G\lr{\Phi} \Phi'\lr{r} A^{(3)}\lr{r}+20 \ell^2 \beta A'\lr{r}\left(\partial_{\Phi}G\lr{\Phi} \Phi'\lr{r} A''\lr{r}+G\lr{\Phi} A^{(3)}\lr{r}\right)+\\
    &+4 \ell^2 \beta G\lr{\Phi} A^{(4)}\lr{r}-36 \ell^2 \beta G\lr{\Phi} A'\lr{r}^4=0\, ,
\end{aligned}
\end{equation}
where primes denote radial derivatives. 

\subsection{UV Asymptotics}

We now analyze the UV limit ($r\to 0$) of the equations of motion \eqref{rr and tt components}-\eqref{rr - tt components}. Our goal is to identify conditions on the potential $G(\Phi)$ such that the UV asymptotics of the vacuum are not modified by the curvature corrections. In particular, we require that the scale factor $A(r)$ follows the same UV expansion as in  \cite{Gursoy:2007er}
\begin{equation}\label{AUV}
    A\lr{r}= -\log\lr{r/\ell}+\frac{4}{9\log\lr{r\Lambda}}+\mathcal{O}\lr{\frac{-\log\lr{\log\lr{r\Lambda}}}{\log\lr{r\Lambda}^2}} \, .
\end{equation} 

Substituting \eqref{AUV} into the equation of motion (\ref{rr and tt components}) gives the following relation at leading order in the $r\to0$ limit
\begin{equation}\label{UVF}
\begin{aligned}
    &3\lr{\log\lr{\Lambda r}}^2\lr{\lr{\log\lr{\Lambda r}}^4-\lr{\log\lr{\Lambda r}}^6r^2\Phi'\lr{r}^2}-4\beta G\lr{\Phi}\lr{\log\lr{\Lambda r}}^6 \\
    &3\beta \partial_{\Phi}G\lr{\Phi}\lr{6\lr{\log\lr{\Lambda r}}^6\Phi'\lr{r}+9\log\lr{\Lambda r}^6r\Phi''\lr{r}}+9\beta \partial_{\Phi}^{2}G\lr{\Phi}\lr{\log\lr{\Lambda r}}^6 + \dots = 0 \, ,
\end{aligned}
\end{equation}
where the dots indicate subleading terms in the $r\to 0$ limit. If $G(\Phi)$ is assumed to go to a constant $G_0$ in the UV limit $(\Phi\to -\infty)$, then \eqref{UVF} determines the leading order behavior of the dilaton to be  
\begin{equation}
\Phi\lr{r}=-\log\lr{-\log\lr{r\Lambda}}\lr{1-\frac{4}{3}\beta G_0} + \dots \, ,
\end{equation}
which is a non-perturbative result in $\beta$. In particular, requiring that the leading order UV behavior of the dilaton is not modified by the curvature corrections imposes that
\begin{equation}
\label{GUV} G(\Phi) \underset{\Phi\to-\infty}{\longrightarrow} 0 \, .
\end{equation}
Requiring that the subleading terms in the UV expansion of the dilaton (which are fixed in ihQCD by matching to the QCD $\beta$-function) are also unchanged, would put further constraints on the way $G$ approaches 0 in the UV. 

The other equation of motion \eqref{rr - tt components} takes the following form at leading order in the UV limit
\begin{equation}\label{UVcV}
\frac{\ell^2 V\lr{\Phi}-12}{r^2}+\frac{8 \beta G\lr{\Phi}}{r^2} = \partial_{\Phi}G\lr{\Phi}+\partial_{\Phi}^{2}G\lr{\Phi}+ \dots \, .
\end{equation}
In the case where $G(\Phi)$ goes to a constant, \eqref{UVcV} implies that the AdS curvature receives a correction proportional to $\beta$  
\begin{equation}
    V\lr{\Phi}\to V_0=\frac{12}{\ell^2}-\frac{8\beta G_0}{\ell^2}.
\end{equation}
This correction to the UV geometry vanishes when \eqref{GUV} is imposed.

\subsection{IR Asymptotics}

We now discuss the IR asymptotics of the vacuum solution, corresponding to $r\to\infty$. The behavior of the background in this limit determines several key qualitative features of the model, including confinement and general properties of the hadronic spectrum. In particular, the best match with QCD properties is obtained when the ihQCD potential $V(\Phi)$ behaves in the IR as   
\begin{equation}\label{VIR}
V(\Phi) \underset{\Phi\to\infty}{\sim} V_{IR}\mathrm{e}^{Q\Phi}\Phi^P \, .     
\end{equation}
The purpose of our analysis is to understand what are the IR asymptotics of $G(\Phi)$ for which these properties hold, once the curvature corrections are introduced. For the remainder of this appendix, $\beta$ will now be assumed to be small, such that the background fields follow the same asymptotics as in \cite{Gursoy:2007cb,Gursoy:2007er}, up to small corrections of order $\mathcal{O}(\beta)$
\begin{equation}
\label{IR asymptotics}
\begin{split}
&A\lr{r}= -Cr^{\alpha}+\dots +\beta \delta A\lr{r} + \mathcal{O}(\beta^2) \, , \quad \alpha>0\, , C>0\\
&\Phi\lr{r}= -\frac{3}{2}A\lr{r}+\frac{3}{4}\log\big|A'\lr{r}\big|+ \dots +\beta \delta\Phi\lr{r} + \mathcal{O}(\beta^2) \, ,
\end{split}
\end{equation}
where 
\begin{equation}
\a = \frac{1}{1-P} \, ,    
\end{equation}
and the dots refer to subleading terms in the $r\to\infty$ limit. We ask how $G\lr{\Phi}$ should behave in the IR such that $\delta A\lr{r}$ and $\delta \Phi\lr{r}$ remain small up to $r\to\infty$.

Substituting \eqref{VIR}-(\ref{IR asymptotics}) into the equations of motion (\ref{rr and tt components})-\eqref{rr - tt components} and keeping only the terms of order $\mathcal{O}(\beta)$, we find the following relation at leading order in $1/r$ 
\begin{equation}
\label{IR EoM}
\begin{aligned}  
&  \frac{8}{3} \alpha^2C^2 \delta \Phi'\lr{r}+ \lr{1-\alpha}r^{-2\alpha}\left(\lr{1-\alpha}\delta A'\lr{r}+ r \delta A''\lr{r}\right) = \\
&\mathrm{e}^{2 C r^\alpha}r^{3(\alpha-1)}\times 6 C^4 \ell^2 \alpha^4 \big(2\alpha C \lr{2G\lr{\Phi} - \partial_{\Phi}G\lr{\Phi}} + \lr{\alpha-1}r^{-\alpha}\partial_{\Phi}^{2}G\lr{\Phi} \big) + \dots + \mathcal{O}(\beta) \, .
\end{aligned}
\end{equation}
\begin{equation}\label{IREOM2}
\begin{aligned}  
&\frac{4}{3}\tilde{V}_{IR} r^{\alpha-1} \delta \Phi\lr{r}+\delta A'\lr{r}\lr{18\alpha C-\lr{\alpha C}^{-1}\tilde{V}_{IR}} -3  r^{1-\alpha} \delta A''\lr{r} = \\
& 18 \ell^2 (\alpha C)^4\, e^{2 C r^\alpha}r^{3 (\alpha-1)}\big(-2G
\lr{\Phi} + \partial_{\Phi}G\lr{\Phi}\big)+ \dots + \mathcal{O}(\beta) \, ,
\end{aligned}
\end{equation}
where we defined
\begin{equation}\label{VIRt}
\tilde{V}_{IR}\equiv V_{IR}\,\lr{3C/2}^P\lr{-\alpha C} \, .
\end{equation}
Note that the right-hand side of \eqref{IR EoM} and \eqref{IREOM2} is multiplied by an exponential that diverges in the $r\to \infty$ limit, whereas the order $\mathcal{O}(\beta^0)$ background has a polynomial behavior (see \eqref{IR asymptotics}). So for the order $\mathcal{O}(\beta)$ corrections to remain sub-leading in the IR, the function $G(\Phi)$ should be such that the exponential divergence cancels in \eqref{IR EoM}. Given the asymptotics of $\Phi$ (\ref{IR asymptotics}), an appropriate ansatz for $G$ is given by
\begin{equation}\label{GIRapp}
G\lr{\Phi}=G_{IR}e^{\gamma\Phi}\Phi^{\delta}.
\end{equation}
which implies the following behavior at leading order in $\beta$ and $1/r$ 
\begin{equation}\label{Gas}
G\lr{\Phi(r)}= G_{IR}\lr{\alpha C}^{\frac{3}{4}\gamma}\lr{\frac{3}{2}C}^{\delta}\mathrm{e}^{\frac{3}{2}\gamma Cr^{\alpha}}r^{\frac{3}{4}\gamma\lr{\alpha-1}+\alpha\delta} + \dots + \mathcal{O}(\beta) \, .
\end{equation}
Plugging this back into \eqref{IR EoM}-\eqref{IREOM2} gives
\begin{equation}
\label{IR EoM2}
\begin{aligned}  
&  \frac{8}{3} \alpha^2C^2 \delta \Phi'\lr{r}+ \lr{1-\alpha}r^{-2\alpha}\left(\lr{1-\alpha}\delta A'\lr{r}+ r \delta A''\lr{r}\right) = \\
&\tilde{G}_{IR}\mathrm{e}^{2Cr^\alpha \lr{1+\frac{3}{4}\gamma}}r^{3(\alpha-1)\lr{1+\frac{1}{4}\gamma}+\alpha\delta} + \dots + \mathcal{O}(\beta) \, ,
\end{aligned}
\end{equation}
\begin{equation}\label{IREOM22}
\begin{aligned}  
&\frac{4}{3}\tilde{V}_{IR} r^{\alpha-1} \delta \Phi\lr{r}+\delta A'\lr{r}\lr{18\alpha C-\lr{\alpha C}^{-1}\tilde{V}_{IR}} -3  r^{1-\alpha} \delta A''\lr{r} = \\
& -\frac{3}{2}(\alpha C)^{-1}\tilde{G}_{IR}\, \mathrm{e}^{2 C r^\alpha\lr{1+\frac{3}{4}\gamma}}r^{3 (\alpha-1)\lr{1+\frac{1}{4}\gamma}+\alpha\delta}+ \dots + \mathcal{O}(\beta) \, ,
\end{aligned}
\end{equation}
where we defined
\begin{equation}\label{defGIRt}
\tilde{G}_{IR} \equiv 12G_{IR}(2-\gamma)\ell^2\lr{\frac{3}{2}C}^\delta \lr{\alpha C}^{5+\frac{3}{4}\gamma} \, .
\end{equation}
From \eqref{IR EoM2}-\eqref{IREOM22}, requiring that the corrections $\d A$ and $\d\Phi$ behave sub-exponentially in the IR imposes that $\gamma$ obeys
\begin{equation}
\gamma\leq-\frac{4}{3}.
\end{equation} 
For the strictly smaller case, the corrections decay exponentially in the $r\to\infty$ limit. The case $\gamma=-\frac{4}{3}$ requires a refined analysis depending on the exponent $\d$, that is carried out below.  

\subsubsection{Special case of $\gamma=-\frac{4}{3}$}

We now consider setting $\gamma=-\frac{4}{3}$, for which the right hand side of \eqref{IR EoM2}-\eqref{IREOM22} behaves polynomially. In this case the potential $G(\Phi)$ is given at leading order by 
\begin{equation}\label{Gas2}
G\lr{\Phi(r)}= G_{IR}\lr{\alpha C}^{-1}\lr{\frac{3}{2}C}^{\delta}\mathrm{e}^{-2 Cr^{\alpha}}r^{-\lr{\alpha-1}+\alpha\delta} + \dots + \mathcal{O}(\beta) \, ,
\end{equation}
and \eqref{IR EoM2}-\eqref{IREOM22} become
\begin{equation}
\label{IR EoM3}
\frac{8}{3} \alpha^2C^2 \delta \Phi'\lr{r}+ \lr{1-\alpha}r^{-2\alpha}\left(\lr{1-\alpha}\delta A'\lr{r}+ r \delta A''\lr{r}\right)= \tilde{G}_{IR}r^{2(\alpha-1)+\alpha\delta} + \dots + \mathcal{O}(\beta) \, ,
\end{equation}
\begin{equation}\label{IREOM23}
\begin{aligned}  
\frac{4}{3}\tilde{V}_{IR} r^{\alpha-1} \delta \Phi\lr{r}&+\delta A'\lr{r}\lr{18\alpha C-\lr{\alpha C}^{-1}\tilde{V}_{IR}} -3  r^{1-\alpha} \delta A''\lr{r} = \\
& -\frac{3}{2}(\alpha C)^{-1}\tilde{G}_{IR}\, r^{2(\alpha-1)+\alpha\delta}+ \dots + \mathcal{O}(\beta) \, .
\end{aligned}
\end{equation}
These equations imply that $\delta A$ and $\delta \Phi$ have a polynomial behavior at $r\to\infty$
\begin{equation}
\label{delta corection asymptotics}
\begin{aligned}  
&\delta A\lr{r}=C_Ar^{\zeta} + \dots + \mathcal{O}(\beta) \, , \\ 
& \delta \Phi\lr{r} = C_{\Phi}r^{\eta} + \dots + \mathcal{O}(\beta) \, ,
\end{aligned}    
\end{equation}
where the exponents are determined in terms of $\delta$ and $\alpha$ as
\begin{equation}
\label{etze} \zeta = \a(3+\d)-1 \quad , \quad\eta=\zeta-\a \, .
\end{equation}
Requiring the corrections to remain small compared with the leading $r^\a$ behavior therefore imposes a constraint on $\delta$
\begin{equation}\label{cd}
\delta < \frac{1}{\a}-2 \, .
\end{equation}

\section{Calculation of shear viscosity}
\label{sec::appB}
In this appendix, we reproduce for completeness the holographic shear viscosity calculation of \cite{Cremonini:2012ny}, which includes four-derivative corrections. We start with the action 
\begin{equation}
\label{Riemannac}
    S=\frac{1}{16 \pi G_5} \int d^5 x \sqrt{-g}\left[R-\frac{4}{3}(\nabla \Phi)^2+V(\Phi)+\ell^2 \beta G(\Phi) R_{\mu \nu \rho \sigma} R^{\mu \nu \rho \sigma}\right]\, ,
\end{equation}
and introduce a shear fluctuation $h_{xy}$ as
\begin{equation}
    g_{\mu\nu}=g_{\mu\nu}^{\lr{0}}+h_{\mu\nu},
\end{equation}
where $g_{\mu\nu}^{\lr{0}}$ is the background metric, of the form 
\begin{equation}
\label{metric}
    ds^2=-a^2\lr{u}dt^2+c^2\lr{u}du^2+b^2\lr{u}d\mathbf{x}^2 \, .
\end{equation}
The background also contains a non-trivial profile for the dilaton, which depends solely on the holographic coordinate $u$
\begin{equation}
\label{dilb} \Phi(x) = \Phi(u) \, .
\end{equation}
Expanding the metric perturbation in Fourier modes
\begin{equation}
    h_{x}^{\, \, y}=\int d^4k \phi_k\lr{u}e^{-i\omega t +ikz}\,,
\end{equation}
and substituting in the action, up to second order in $\phi_k$ we obtain 
\begin{equation}
\label{effective action}
\begin{split}
S_{e f f} = \int \frac{d^4 k}{(2 \pi)^4} d u & {\left[A(u) \phi_k^{\prime \prime} \phi_{-k}+B(u) \phi_k^{\prime} \phi_{-k}^{\prime}+C(u) \phi_k^{\prime} \phi_{-k}\right.} \\
& \left.+D(u) \phi_k \phi_{-k}+E(u) \phi_k^{\prime \prime} \phi_{-k}^{\prime \prime}+F(u) \phi_k^{\prime \prime} \phi_{-k}^{\prime}\right]\, ,
\end{split}
\end{equation}
where primes denote $u$ derivatives. From the Kubo relation, one arrives at the shear viscosity in terms of the horizon data \cite{Buchel:2004di,Myers:2009ij}
\begin{equation}
\label{shear visc formula}
    \eta=\left.\frac{1}{8 \pi G_5}\left[\sqrt{-\frac{g_{u u}}{g_{t t}}}\left(A-B+\frac{F^{\prime}}{2}\right)+\left(E\left(\sqrt{-\frac{g_{u u}}{g_{t t}}}\right)^{\prime}\right)^{\prime}\right]\right|_{u=u_h}\, ,
\end{equation}
where the whole expression is evaluated at the horizon. 

The various coefficients that appear in the effective action (\ref{effective action}) are given by
\begin{equation}
    \begin{split}
         & A\lr{u}=\frac{2 a\lr{u}b^2\lr{u}\left(b\lr{u}c^3\lr{u}+4\ell^2 \beta G\lr{\Phi}\left(b^{\prime}\lr{u} c^{\prime}\lr{u}-c\lr{u} b^{\prime \prime}\lr{u}\right)\right)}{c\lr{u}^4}\,, \\ \\
         &B\lr{u}= \frac{b\lr{u}}{2 a\lr{u} c^5\lr{u}}\big[ 4\ell^2\beta b^2\lr{u}c^2\lr{u}G\lr{\Phi}\lr{a^{\prime}\lr{u}}^2+a^2\lr{u}\big(16\ell^2\beta c^2\lr{u}G\lr{\Phi}\lr{b^{\prime}\lr{u}}^2 \\ 
         &+\!b^2\lr{u}\lr{3c^4\lr{u}+4\ell^2\beta G\lr{\Phi}\lr{c^{\prime}\lr{u}}^2}\!\!-8\ell^2\beta b\lr{u}c\lr{u}G\lr{\Phi}\lr{b^{\prime}\lr{u}c^{\prime}\lr{u}+c\lr{u}b^{\prime\prime}\lr{u}}\!\big)\big], \\ \\
         &C\lr{u}=-\frac{1}{c^5\lr{u}a\lr{u}}\big[ -a\lr{u}b^3\lr{u}c^4\lr{u}a^{\prime}\lr{u}+4\ell\beta b^2\lr{u}c^2\lr{u}G\lr{\Phi}\lr{a^{\prime}\lr{u}}^2b^{\prime}\lr{u}+\\
         &+a^2\lr{u}\big(8\ell^2\beta c^2\lr{u}G\lr{\phi}\lr{b^{\prime}}^3+b^3\lr{u}c^3\lr{u}c^{\prime}\lr{u}-\\
         &-8\ell^2\beta b\lr{u}c\lr{u}G\lr{\Phi}b^{\prime}\lr{u}\lr{b^{\prime}\lr{u}c^{\prime}\lr{u}-c\lr{u}b^{\prime\prime}\lr{u}}-\\
         &-4b^2\lr{u}\lr{c^4\lr{u}b^{\prime}\lr{u}-\ell^2\beta G\lr{\Phi}b^{\prime}\lr{u}\lr{c^{\prime}\lr{u}}^2+\ell^2\beta c\lr{u}G\lr{\Phi}c^{\prime}\lr{u}b^{\prime\prime}\lr{u}}\big)\big]\,,\\ \\
         &D\lr{u} = \frac{1}{2c^5\lr{u}a\lr{u}b\lr{u}}\Big[2a\lr{u}b^3\lr{u}c^3\lr{u}\times\\
         &\times\left(-b\lr{u} a^{\prime}\lr{u}c^{\prime}\lr{u}+c\lr{u}\left(3 a^{\prime}\lr{u}b^{\prime}\lr{u}+b\lr{u} a^{\prime\prime}\lr{u}\right)\right)- \\ 
         & -4 \ell^2 \beta b^2\lr{u} G\lr{\Phi}\left(b^2\lr{u}\lr{a^{\prime}\lr{u}}^2 \lr{c^{\prime}\lr{u}}^2-2 b^2\lr{u} c\lr{u} a^{\prime}\lr{u} c^{\prime}\lr{u} a^{\prime\prime}\lr{u}+\right.\\
         &\left.+c^2\lr{u}\left(3 \lr{a^{\prime}\lr{u}}^2 \lr{b^{\prime}\lr{u}}^2+b^2\lr{u} \lr{a^{\prime \prime}\lr{u}}^2\right)\right)+\\
         &+a^2\lr{u}\Big(b^4\lr{u} c^6\lr{u}\left(2 \partial_u\Phi\partial^u\Phi-V\lr{\Phi}\right)-12 \ell^2 \beta c^2\lr{u} G\lr{\Phi} \lr{b^{\prime}(u)}^4 + \\ 
         &+6 b^3\lr{u} c^3\lr{u}\left(-b^{\prime}\lr{u}c^{\prime}\lr{u}+c\lr{u} b^{\prime \prime}\lr{u}\right)+\\
         &+6 b^2\lr{u}\big(c^4\lr{u} \lr{b^{\prime}\lr{u}}^2-2 \ell^2 \beta G\lr{\Phi} \lr{b^{\prime}\lr{u}}^2 \lr{c^{\prime}\lr{u}}^2+ \\ 
         &+4 \ell^2 \beta c\lr{u} G\lr{\Phi} b^{\prime}\lr{u} c^{\prime}\lr{u} b^{\prime \prime}\lr{u}-2 \ell^2 \beta c^2\lr{u} G\lr{\Phi} \lr{b^{\prime \prime}\lr{u}}^2 \big)\Big)\Big]\,,\\ \\
         &E\lr{u}=\frac{2 \ell^2 \beta a\lr{u} b^3\lr{u}G\lr{\Phi}}{c^3\lr{u}} \, , \\\\
         & F\lr{u}=\frac{4 \ell^2 \beta a\lr{u}b^2\lr{u} G\lr{\Phi}\left(2 c\lr{u}b^{\prime}\lr{u}-b\lr{u}c^{\prime}\lr{u}\right)}{c^4\lr{u}} \, ,
    \end{split}
\end{equation}
where we have already applied the relevant limit in the Kubo relation, $\omega\to0$ and $k=0$.

For $\eta/s$ we also need to compute the entropy density from horizon data. In the presence of curvature corrections, the black brane entropy $S$ is computed from Wald's formula as
\begin{equation}
\label{Wald's formula}
S=-2 \pi \oint_{\Sigma} d^3 x \sqrt{-h} \frac{\delta \mathcal{L}}{\delta R_{\mu \nu \rho \sigma}} \epsilon_{\mu \nu} \epsilon_{\rho \sigma},    
\end{equation}
where $\Sigma$ refers to the horizon, with induced metric $h$. The tensor $\epsilon_{\mu \nu}$ is binormal to $\Sigma$ and is normalized as $\epsilon_{\mu \nu}\epsilon^{\mu \nu}=-2$ and $\epsilon_{\mu \nu}=-\epsilon_{\nu \mu}$. Using that
\begin{equation}
    \frac{\delta \mathcal{L}}{\delta R_{\mu \nu \rho \sigma}}=\frac{1}{16 \pi G_5} \left[g^{\nu\sigma}g^{\rho\mu}+2\ell^2 \beta G(\Phi)R^{\mu \nu \rho \sigma}\right]\, , 
\end{equation}
and expanding the background (\ref{metric})-\eqref{dilb} near the horizon as 
\begin{equation}
\label{metric expansions}
\begin{split}
&a(u)^2  =a_0(1-u)+a_1(1-u)^2+a_2(1-u)^3+\ldots \\
&b(u)^2  =b_0(1+(1-u)+\ldots) \\
&c(u)^2 =c_0(1-u)^{-1}+c_1+c_2(1-u)+\ldots \\
&\Phi(u) =\Phi_h+\varphi_1(1-u)+\varphi_2(1-u)^2+\ldots
\end{split}
\end{equation}
we find
\begin{equation}
S=\frac{b_{0}^{3/2}}{4G_{5}} V_{3}\left[1-\frac{3a_1c_0-a_0c_1}{a_0c_{0}^{2}}\ell^2 \beta G(\Phi_{h})\right]\, ,   
\end{equation}
with $V_3$ the three-dimensional spatial volume. The entropy density $s$ is thus given by 
\begin{equation}
\label{Rieman entropy}
 s=\frac{b_{0}^{3/2}}{4G_{5}}\left[1-\frac{3a_1c_0-a_0c_1}{a_0c_{0}^{2}}\ell^2 \beta G(\Phi_{h})\right]\, .
\end{equation}

Employing (\ref{metric expansions}) in (\ref{shear visc formula}) and dividing by (\ref{Rieman entropy}) we finally arrive at
\cite{Cremonini:2011ej,Cremonini:2012ny}
\begin{equation}
\label{eta over s}
    \frac{\eta}{s}=\frac{1}{4\pi}\left[1-\frac{\beta\ell^2}{c_0}\lr{G\lr{\Phi_h}+2\sqrt{\frac{3}{2}}\varphi_1\partial_{\Phi}G\lr{\Phi_h}}\right]\, .
\end{equation}

This result can be fully expressed in terms of the potentials $V$ and $G$ at the horizon. To see this, we first change coordinates to 
\begin{equation}
ds^2=f^{-1}\lr{r}dr^2+e^{2A\lr{r}}\lr{d\mathbf{x}^2-f\lr{r}dt^2},\quad \Phi=\Phi\lr{r}\, ,
\end{equation}
with near-horizon expansions  
\begin{equation}
    \begin{split}
         A(r) & =A_h+A_1\left(r-r_h\right)+\cdots \\ f(r) & =f_1\left(r-r_h\right)+\cdots \\ \Phi(r) & =\Phi_h+\Phi_1\left(r-r_h\right)+\cdots
    \end{split}
\end{equation}
The relation to (\ref{metric expansions}) is given by
\begin{equation}
    \varphi_1=\frac{1}{2 A_1} \Phi_1, \quad c_0=\frac{1}{2 f_1 A_1} \, .
\end{equation}
Substituting back into (\ref{eta over s}) we find
\begin{equation}
\label{eta over s as A1,f1}
    \frac{\eta}{s}=\frac{1}{4\pi}\left[1-2f_1 A_1\beta\ell^2\lr{G\lr{\Phi_h}+\sqrt{\frac{2}{3}}\frac{\Phi_1}{A_1}\partial_{\Phi}G\lr{\Phi_h}}\right].
\end{equation}

The metric equations can then be conveniently expressed as \cite{Gursoy:2008za}  
\begin{equation}
\begin{split}
    \frac{dA}{dr}&=A_1=-\frac{C}{\ell} \frac{S^{\frac{1}{3}}}{T} V(\Phi_h) \\
    \frac{d\Phi}{dr}&=\Phi_1 =\frac{3 C}{4 \ell} \frac{S^{\frac{1}{3}}}{T} \sqrt{\frac{3}{2}}\partial_{\Phi}V(\Phi_h) \\
    f_1 &=-M_p(4 \pi)^{\frac{4}{3}} \frac{T}{S^{\frac{1}{3}}}\,,
\end{split}
\end{equation}
where $S,T$ are the entropy and temperature and $C$ is a constant that is unimportant for our purposes. Substituting $A_1$, $\Phi_1$ and $f_1$ back into our expression for $\eta/s$ (\ref{eta over s as A1,f1}) we obtain (\ref{eta over s theory}) which was first derived in \cite{Cremonini:2012ny}.
\section{Invariance of $\eta/s$ under modification of curvature terms}
\label{sec::appC}
In this work, we considered a single type of curvature squared correction to the bulk Lagrangian, corresponding to the Riemann squared $R_{\m\n\r\s}^2$. The most general action at this order in curvature however contains more terms and takes the form
\begin{equation}\label{Sgen}
S=\frac{1}{16 \pi G_5} \int\! d^5 x \sqrt{-g}\left[R-\frac{4}{3}(\nabla \Phi)^2+V(\Phi)+\ell^2 \beta\lr{\lambda_1R^2\!+\!\lambda_2R_{\mu\nu}R^{\mu\nu} \!+\! \lambda_3 R_{\mu \nu \rho \sigma} R^{\mu \nu \rho \sigma}}\right] ,  
\end{equation}
where all the $\lambda_i$ may be taken to depend on the dilaton $\Phi$. In this appendix, we show that the shear viscosity per entropy $\eta/s$ is actually independent of $\l_1$ and $\l_2$, which implies that \eqref{eta over s theory} is actually the most general expression for $\eta/s$ at quadratic order in curvature. 

As a warm-up, we start by reviewing the result of \cite{Kats:2007mq}, which applies to constant coefficients $\lambda_i$. Following the same procedure as in appendix \ref{sec::appB}, we find 
\begin{equation}
\begin{aligned}
 \eta= \frac{b_0^{3/2}}{16\pi G}\bigg[1-\ell^2\beta\bigg(\frac{6a_0c_0+3a_1c_0-a_0c_1}{a_0c_0^2}\lambda_1&+\frac{3\lr{a_0+a_1}c_0-a_0c_1}{2a_0c_0^2}\lambda_2+\\
 &+\frac{a_0c_0+3a_1c_0-a_0c_1}{a_0c_0^2}\lambda_3\bigg)\bigg]\, ,
\end{aligned}
\end{equation}
and
\begin{equation}
    s=\frac{b_0^{3/2}}{4G}\left[1-\ell^2\beta\lr{\frac{6a_0c_0+3a_1c_0-a_0c_1}{a_0c_0^2}\lambda_1\!+\!\frac{3\lr{a_0+a_1}c_0-a_0c_1}{2a_0c_0^2}\lambda_2\!+\!\frac{3a_1c_0-a_0c_1}{a_0c_0^2}\lambda_3}\right]\!.
\end{equation}
Up to order $\mathcal{O}(\beta)$, the ratio $\eta/s$ is then given by\footnote{We again refer to appendix \ref{sec::appB} for the definitions of the various coefficients in the near-horizon expansion of the background.}
\begin{equation}
\begin{aligned}
  \frac{\eta}{s}&=\frac{1}{4\pi}\left[1-\frac{\beta\ell^2}{c_0}\lambda_3\right]+\mathcal{O}(\beta^2)\\ 
  &= \frac{1}{4\pi}\left[1-2f_1A_1\beta\ell^2\lambda_3\right]+\mathcal{O}(\beta^2)\\
  &=\frac{1}{4\pi}\left[1-\frac{2}{3}\beta\ell^2\lambda_3V\lr{\Phi_h}\right]+\mathcal{O}(\beta^2) \, .
\end{aligned}
\end{equation}
This shows that $\eta/s$ is indeed independent of $\lambda_1$ and $\lambda_2$.

This property of the shear viscosity per entropy can be explained by considering  redefinitions of the metric of the form
\begin{equation}
    g_{\mu\nu}\to g_{\mu\nu}+ag_{\mu\nu}R+bR_{\mu\nu}.
\end{equation}
with $a$ and $b$ two constants of order $\mathcal{O}\lr{\beta}$. Under such a redefinition, the action \eqref{Sgen} is brought to the following form \cite{Kats:2007mq} 
\begin{equation}
S=\frac{1}{16 \pi G_5}\! \int\! d^5 x \sqrt{-g}\left[\frac{R}{\kappa}-\frac{4}{3}(\nabla \Phi)^2+V(\Phi)+\ell^2\beta\lr{\tilde{\lambda}_1R^2\!+\!\tilde{\lambda}_2R_{\mu\nu}R^{\mu\nu}\!+\! \lambda_3 R_{\mu \nu \rho \sigma} R^{\mu \nu \rho \sigma}}\right] \!\!,   
\end{equation}
where
\begin{equation}
    \frac{1}{\kappa}=1+ \frac{5a+b}{2}\lr{V\lr{\Phi}-\frac{4}{3}(\nabla \Phi)^2}\, , \quad \Tilde{\lambda}_1=\lambda_1+\frac{3a+b}{2}\, ,\quad \Tilde{\lambda}_2=\lambda_2-b.
\end{equation}
By choosing $b=\lambda_2$ and $a=\frac{-2\lambda_1-\lambda_2}{3}$, we are therefore left with an action containing only the Riemann squared term at the quadratic level, albeit with a rescaled Ricci scalar. This rescaling will not produce a deviation from the universal result of $1/4\pi$ since $\eta/s$ does not depend on $\kappa$. This is because in formula (\ref{shear visc formula}) the change of $\kappa$ only amounts to a rescaling of the $A$,$B,E$ and $F$ coefficients
\begin{equation}
\begin{split}
& A\to A\kappa^{-1}+\mathcal{O}(\beta^2) \quad,\quad B\to B\kappa^{-1}+\mathcal{O}(\beta^2)\, ,\\
& E \to E + \mathcal{O}(\beta^2) \quad,\quad F\to F+\mathcal{O}(\beta^2) \, .
\end{split}
\end{equation}
This implies a shift in the shear viscosity 
\begin{equation}\label{teta}
\eta \to \eta + \frac{b_0^{3/2}}{16\pi G}(\kappa^{-1}-1) + \mathcal{O}(\beta^2) \, .
\end{equation}
The entropy density however receives the same shift
\begin{equation}\label{ts}
s \to s + \frac{b_0^{3/2}}{16\pi G}(\kappa^{-1}-1) + \mathcal{O}(\beta^2) \, ,
\end{equation}
so that the ratio $\eta/s$ is invariant under the rescaling by $\kappa$ up to order $\mathcal{O}(\beta)$.

We now turn to the case where the higher curvature couplings are functions of the dilaton $\lambda_i(\Phi)$. The general formula for the shear viscosity and entropy density with the action \eqref{Sgen} can still be expressed in terms of horizon data through \eqref{shear visc formula}. The difference with constant coefficients can therefore only arise if $\eta$ and $s$ depend on derivatives of $\l_i(\Phi)$ at the horizon. It can be checked however that derivatives of $\lambda_1(\Phi)$ and $\lambda_2(\Phi)$ appear in the expressions of neither $\eta$ nor $s$, so that $\eta/s$ should remain independent of $\lambda_1$ and $\lambda_2$ even with dilaton dependence. This indicates that the metric redefinition is still valid even if the coefficients are not constant.

\section{Calculation of the bulk viscosity}
\label{sec::appD}
We detail here the derivation of the expression for the bulk viscosity \eqref{zetas1}, which generalizes the calculation of \cite{Buchel:2023fst} to the case where the coefficients of the curvature corrections depend on the dilaton. We start from a general ansatz for the background metric and dilaton  
\begin{equation}
ds^2=-c_1\lr{r}^2dt^2+c_2^2\lr{r}d\mathbf{x}^2+c_3^2\lr{r}dr^2, \quad \Phi=\Phi(r)\, .
\end{equation}
Substituting this ansatz in the equations of motion presented in Appendix \ref{sec::bg} gives the equations obeyed by the ansatz fields
\begin{equation}
\begin{split}
&\Phi^{\prime\prime}+\frac{c_1^{\prime}\Phi^{\prime}}{c_1}+\frac{3c_2^{\prime}\Phi^{\prime}}{c_2}-\frac{c_3^{\prime}\Phi^{\prime}}{c_3}+\frac{3}{8}c_3^2\partial_{\Phi}V +\mathcal{O}(\beta) = 0\,, \\
&\lr{\Phi^{\prime}}^2+\frac{3}{4}c_3^2V-\frac{9c_1^{\prime}c_2^{\prime}}{2c_1c_2}-\frac{9(c_2^{\prime})^2}{2c_2^2}+\mathcal{O}(\beta)=0\,,\\ &c_2^{\prime\prime}+\frac{\lr{c_2^{\prime}}^2}{c_2}-\frac{c_2^{\prime}c_3^{\prime}}{c_3}+\frac{2}{9}c_2\lr{\Phi^{\prime}}^2-\frac{1}{6}c_2c_3^2V+\mathcal{O}(\beta)=0\,,\\ &c_1^{\prime\prime}+\frac{2c_1^{\prime}c_2^{\prime}}{c_2}-\frac{c_1^{\prime}c_3^{\prime}}{c_3}+\frac{c_1(c_2^{\prime})^2}{c_2^2}-\frac{1}{6}c_1c_3^2V+\frac{2}{9}c_1\lr{\Phi^{\prime}}^2+\mathcal{O}(\beta)=0 \, ,
\end{split}
\end{equation}
where the terms of order $\mathcal{O}(\beta)$ were not written explicitly to avoid clutter. 

To compute the relevant retarded Green's function (see section \ref{sec::viscos}) we expand the metric and dilaton, considering SO(3) invariant perturbations\footnote{Our definition for $h_{tt}$ differs from \cite{Buchel:2023fst} by a minus sign.} in the axial gauge $h_{rr}=h_{rt}=0$
\begin{equation}
\begin{split}
&\d c_1\lr{r,t} = \frac{ h_{tt}\lr{r,t}}{2c_1\lr{r}} \, , \\
&\d c_2\lr{r,t} = \frac{ h_{11}\lr{r,t}}{2c_2\lr{r}} \, ,\\
&\d c_3\lr{r,t} = 0 \, , \\
&\d\Phi\lr{r,t} = \Psi\lr{r,t}.
\end{split} 
\end{equation}
It is convenient to introduce new variables $H_{11},H_{tt}$ such that
\begin{equation}
h_{tt}\lr{r,t} \equiv e^{-i\omega t}c_1^2H_{00}\lr{r}  \, , \quad h_{11}\lr{r,t} \equiv e^{-i\omega t}c_2^2H_{11}\lr{r} \, ,
\end{equation}
and the gauge invariant scalar fluctuation $Z$ defined by
\begin{equation}\label{defZ}
\Psi\lr{r,t} \equiv e^{-i\omega t}\lr{Z\lr{r}+\frac{\Phi^{\prime}\lr{r}c_2\lr{r}}{2c_2^{\prime}\lr{r}}H_{11}\lr{r}}\, .
\end{equation}
The linearized equations of motion give the equations obeyed by the fluctuations
\begin{equation}
    H^{\prime}+\frac{8c_2c_3\Phi^{\prime}}{9c_2^{\prime}}Z+ \mathcal{O}(\beta) = 0 \, , \quad H\equiv\frac{c_2c_3}{c_2'}H_{11},
\end{equation}
\begin{equation}
\begin{split}
    &H_{00}^{\prime}-\frac{8c_2\Phi^{\prime}}{9c_2^{\prime}}Z^{\prime}+\frac{c_2}{27c_1(c_2^{\prime})^2}\lr{24c_1^{\prime}c_2^{\prime}\Phi^{\prime}-c_1c_3^{2}\lr{9c_2^{\prime}\partial_{\Phi}V+8c_2V\Phi^{\prime}}}Z\\
    &\qquad\qquad\quad\quad+\lr{\frac{\omega^2c_3}{c_1^2}-\frac{1}{3}c_3V+\frac{(c_1^{\prime})^2}{c_3c_1^2}+\frac{3c_1^{\prime}c_2^{\prime}}{c_1c_2c_3}}H+\mathcal{O}(\beta)=0,
    \end{split}
\end{equation}
where we made use of the background equations of motion to simplify the above expressions. As before, we only provide the zeroth order results as the order $\mathcal{O}(\beta)$ expressions are too cumbersome to present here. Remarkably, the equation for $Z$ decouples from the other fluctuations once the constraints above are used in tandem with the background equations. The resulting equation for $Z$ takes the form
\begin{equation}\label{EZ}
Z^{\prime\prime}\!+\!\lr{\frac{c_1^{}\prime}{c_1}\!+\!\frac{3c_2^{\prime}}{c_2}\!-\!\frac{c_3^{\prime}}{c_3}}\!Z^{\prime} \!+c_3^2\!\lr{\frac{\omega^2}{c_1^2}\!+\!\frac{4}{3}V\!-\!\frac{2c_2^2c_3^2V^2}{9(c_2^{\prime})^2}\!+\!\frac{4c_2Vc_1^{\prime}}{3c_1c_2^{\prime}}\!+\!\frac{2c_2\partial_{\Phi}V\Phi^{\prime}}{3c_2^{\prime}}\!+\!\frac{3}{8}V^{\prime\prime}\!}\!Z +\mathcal{O}(\beta) \!=\! 0 ,
\end{equation}
where the $\mathcal{O}(\beta)$ terms are also only functions of $Z$. 

The bulk viscosity can be computed from the effective action at quadratic order in perturbations\footnote{The action \eqref{Sc} is actually the \textit{complexified} version of the quadratic on-shell action; see \cite{Buchel:2023fst} for details.} 
\begin{equation}\label{Sc}
    S_c=\frac{1}{16\pi G_5}\int dr\mathcal{L}_c\{h_{11,\omega},h_{00,\omega},p_{\omega},h_{11,\omega}^{*},h_{00,\omega}^{*},p_{\omega}^{*}\},
\end{equation} 
where we defined
\begin{equation}\label{mf}
    h_{t t}(t, r)=e^{-i \omega t} h_{00,\omega}\lr{r}, \quad h_{11}(t, r)=e^{-i \omega t} h_{11,\omega}\lr{r}, \quad \psi(t, r)=e^{-i \omega t} p_{\omega}\lr{r}.
\end{equation} 
Since the expression for the effective action is very lengthy and does not add to the understanding of the derivation, we do not present it here and refer to \cite{Buchel:2023fst} for more details. In the following we suppress the $\omega$ indices and instead use ``$*$" to refer to conjugate fields. The effective action is invariant under a U(1) symmetry which acts on the mode functions in \eqref{mf} by a simultaneous phase rotation 
\begin{equation}
h_{00,\omega} \to \mathrm{e}^{i\alpha}h_{00,\omega} \quad,\quad h_{11,\omega} \to \mathrm{e}^{i\alpha}h_{11,\omega} \quad,\quad p_{\omega} \to \mathrm{e}^{i\alpha}p_{\omega} \, .
\end{equation}
This symmetry is associated with the conserved current 
\begin{equation}
J_{N}=-i\lr{J_{\omega}-J_{-\omega}}=2 \text{Im}J_{\omega}.
\end{equation}
with
\begin{equation}
\begin{split}
    J_{\omega}&=\frac{\partial \mathcal{L}}{\partial h_{11}^{* \prime}}\delta h_{11}^{*}+\frac{\partial \mathcal{L}}{\partial h_{11}^{* \prime\prime}}\delta h_{11}^{* \prime}-\lr{\frac{\partial \mathcal{L}}{\partial h_{11}^{*\prime\prime}}}^{\prime} \delta h_{11}^{*} \\
   &+\frac{\partial \mathcal{L}}{\partial h_{00}^{* \prime}}\delta h_{00}^{*}+\frac{\partial \mathcal{L}}{\partial h_{00}^{* \prime\prime}}\delta h_{00}^{* \prime}-\lr{\frac{\partial \mathcal{L}}{\partial h_{00}^{*\prime\prime}}}^{\prime} \delta h_{00}^{*} \\
   &+\frac{\partial \mathcal{L}}{\partial p^{* \prime}}\delta p^{*}+\frac{\partial \mathcal{L}}{\partial p^{* \prime\prime}}\delta p^{* \prime}-\lr{\frac{\partial \mathcal{L}}{\partial p^{*\prime\prime}}}^{\prime} \delta p^{*} .
\end{split}
\end{equation}
After some lengthy but straightforward calculations, we find:
\begin{equation}\label{Jc}
\begin{split}
    J_{\omega}= & \frac{1}{12 c_1^4 c_2^2 c_3} \Bigg[9 c_2^5 h_{00}^{*} h_{00} c_1^{\prime}-9 c_1^2 c_2^3 h_{00}^{*} h_{11} c_1^{\prime}-3c_1 c_2^5( h_{00} h_{00}^{*\prime}+h_{00}^{*}h_{00}^{\prime})+ 
    \\ & c_1^5\left(9 c_2 h_{11} h_{11}^{*\prime}+9 h_{11}^{*}\left(h_{11} c_2^{\prime}-c_2 h_{11}^{\prime}\right)+16 c_2^5 p^{*} p^{\prime}+24 c_2^3 p^{*} h_{11} \Phi^{\prime}\right)+ 
    \\ & c_1^3 c_2^2\left(-9 h_{11}^{*} h_{00}c_2^{\prime}+c_2\left(9h_{00} h_{11}^{*\prime}+9 h_{11}h_{00}^{*\prime}+8 c_2^2 p^{*}h_{00}\Phi^{\prime}\right)\right)\Bigg] + \mathcal{O}(\beta) \, .
\end{split}
\end{equation}

For our purpose, we wish to compute the current \eqref{Jc} at leading order in the hydrodynamic approximation. Assuming  $\mathfrak{w}=\frac{\omega}{2\pi T}$ to be much smaller than 1, the fluctuations can be expanded up to order $\mathcal{O}(\mathfrak{w})$ as
\begin{equation}
\label{defz0z1}
    Z=\left(\frac{c_1}{c_2}\right)^{-i \mathfrak{w}}\left(z_{0}+i \mathfrak{w} z_{1}\right) + \mathcal{O}(\mathfrak{w}^2) \quad,\quad H=H_0+i \mathfrak{w} H_1+\mathcal{O}(\mathfrak{w}^2) \, ,
\end{equation}
\begin{equation}
H_{00}=H_{00,0}+i \mathfrak{w} H_{00,1}+\mathcal{O}(\mathfrak{w}^2) \, .
\end{equation}

To extract the bulk viscosity using the Kubo relation, we need the imaginary part of the current at order $\mathcal{O}(\mathfrak{w})$. After making use of the equations of motion for $H,H_{00}$ we find that the expression simplifies significantly and the conserved current only depends on $z_0,z_1$. Its imaginary part takes the form
\begin{equation}
\mathrm{Im}J_{\omega}=-\mathfrak{w}J_{1} \, ,
\end{equation}
with
\begin{equation}
\begin{split}
  J_1= & \frac{4 c_2^2\left(c_1 z_{0}^{2} c_{2}^{\prime}-c_2\left(z_{0}^{2} c_1^{\prime}+c_1 z_1 z_0^{\prime}-c_1 z_0 z_1^{\prime}\right)\right)}{3c_3} \\ +&\frac{16 \ell^2 \beta}{27 c_1^2 c_3^3 c_2^{\prime 2}} \left(c_1 z_0^2 c_2^{\prime}-c_2\left(z_0^2 c_1^{\prime}+c_1 z_1 z_0^{\prime}-c_1 z_0 z_1^{\prime}\right)\right) \Big[108 c_2^2 G\lr{\Phi} c_1^{\prime 2} c_2^{\prime 2}- \\ & 6 c_1 c_2 c_1^{\prime} c_2^{\prime}\left(4 c_2^2 c_3^2 G\lr{\Phi} V\lr{\Phi}-21 G\lr{\Phi} c_2^{\prime 2}+6 c_2 c_2^{\prime} \partial_{\Phi}G\lr{\Phi} \Phi^{\prime}\right)+ \\ & c_1^2\left(c_2^4 c_3^4 G\lr{\Phi} V\lr{\Phi}^2-9 c_2^2 c_3^2 G\lr{\Phi} V\lr{\Phi} c_2^{\prime 2}+18 G\lr{\Phi} c_2^{\prime 4}-\right. \\ & \left.18 c_2 c_2^{\prime 3} \partial_{\Phi}G\lr{\Phi}\Phi^{\prime}+c_2^3 c_3^2 c_2^{\prime}\left(5 V\lr{\Phi} \partial_{\Phi}G\lr{\Phi}+2 G\lr{\Phi} \partial_{\Phi}V\lr{\Phi}\right) \Phi^{\prime}\right)\Big] .
  \end{split}
\end{equation}
A key observation in this calculation is that since the current is radially conserved one may evaluate it at any point in the holographic direction and in particular evaluate it at the horizon. This simplifies the expression and gives $\zeta$ only in terms of horizon data. Expanding around the horizon as in (\ref{metric expansions}):
\begin{equation}
\begin{split}
&c_1^2\lr{u}  =a_0(1-u)+a_1(1-u)^2+a_2(1-u)^3+\ldots \\
&c_2^2\lr{u}  =b_0(1+(1-u)+\ldots) \\
&c_3^2\lr{u} =c_0(1-u)^{-1}+c_1+c_2(1-u)+\ldots \\
&\Phi\lr{u} =\Phi_h+\varphi_1(1-u)+\varphi_2(1-u)^2+\ldots
\end{split}
\end{equation}
Regularity at the horizon yields the following constraints:
\begin{equation}
    c_0=\frac{3}{2V_h}\, , \quad \varphi_1=-\frac{9\partial_{\Phi}V_{h}}{16V_h}\, , \quad 
    c_1=  \frac{15}{4V_h}+\frac{81\lr{\partial_{\Phi}V_h}^2}{128V^{3}_{h}}\, , \quad a_1=a_0\lr{\frac{1}{2}+\frac{9(\partial_{\Phi}V_h)^2}{64V_h^2}},
\end{equation}
Expanding the current at the horizon and making use of the constraints, we find 
\begin{equation}
    J_{1h}=\frac{2}{3}\sqrt{\frac{a_0}{c_0}}b_0^{3/2}z_0^2\lr{1+\frac{\ell^2\beta}{2} \partial_{\Phi}V\lr{\partial_{\Phi}G+\frac{G\partial_{\Phi}V}{V}}},
\end{equation}
\\
where the entire expression is evaluated at the horizon. The  Kubo relation
\begin{equation}
    \zeta=-\frac{4}{9}\lim_{\omega\to 0}\frac{1}{\omega}\text{Im}G_R=-\frac{1}{18 \pi G_5}\lim_{\omega\to 0}\frac{1}{\omega}\text{Im}J_{\omega}
\end{equation}
then yields
\begin{equation}
    \zeta=\frac{\mathfrak{w}}{27\pi G_5}\sqrt{\frac{a_0}{c_0}}b_0^{3/2}z_0^2\lr{1+\frac{\ell^2\beta}{2} \partial_{\Phi}V\lr{\partial_{\Phi}G+\frac{G\partial_{\Phi}V}{V}}} \, ,
\end{equation}
which is valid for a normalization of the metric fluctuations such that 
\begin{equation}\label{H11UV}
H_{11}(r=0) = 1 \, .
\end{equation}
Using (\ref{Rieman entropy}) results in the following for the bulk viscosity per entropy
\begin{equation}
 s=\frac{b_{0}^{3/2}}{4G_{5}}\left[1-\frac{3a_1c_0-a_0c_1}{a_0c_{0}^{2}}\ell^2 \beta G\right]=\frac{b_{0}^{3/2}}{4G_{5}}\left[1+\frac{2\ell^2 \beta}{3} G V\right].
\end{equation}
Taking the ratio between the two expressions and keeping terms up to order $\mathcal{O}(\beta)$ we finally arrive at
\begin{equation}
    \frac{\zeta}{s}=\frac{8z_0^2}{27\pi}\left[1+\frac{1}{2}\ell^2\beta\lr{-\frac{4}{3}GV+\frac{G(\partial_{\Phi}V)^2}{V}+\partial_{\Phi}G\partial_{\Phi}V}\right]\, .
\end{equation}
One can easily generalize this to (\ref{zetas2}) in the presence of all 4-derivative corrections. 
\section{Numerical computation of the bulk viscosity}
\label{sec::appE}
We outline here our procedure for the numerical evaluation of $\zeta/s$ and to fit the Bayesian data. We find it more convenient for the numerical calculation to change coordinates as follows
\begin{equation}
    c_1\to\sqrt{f\lr{r}}e^{A\lr{r}}\, , \quad c_2\to e^{A\lr{r}}\, , \quad c_3\to \frac{e^{A\lr{r}}}{\sqrt{f\lr{r}}}\,.
\end{equation}
In addition, we change coordinates to use the scale factor $A$ instead of the radial coordinate $r$, for which it is useful to introduce the variable
\begin{equation}
    q\lr{A}\equiv\frac{e^A\lr{r}}{A^{\prime}\lr{r}}\, .
\end{equation}
The first step to compute the fluctuation is to solve the background equations at zeroth order in $\beta$, where the ansatz functions are expanded as
\begin{equation}\label{expbg}
f  =f_0+\beta f_1+\mathcal{O}(\beta^2)\, , \quad q =q_0+\beta q_1+\mathcal{O}(\beta^2)\,,\quad
\lambda = \lambda_0 +\beta \lambda_1+\mathcal{O}(\beta^2)\,,
\end{equation}
with $\l$ the exponential of the dilaton
\begin{equation}
\l\equiv \mathrm{e}^\Phi \, ,    
\end{equation}
which is a better variable for the numerical analysis. At leading order, the three equations that we solve are
\begin{equation}
\label{numerical equations}
\begin{split}
    &\frac{f_{0}^{\prime\prime}\lr{A}}{\sqrt{f_{0}\lr{A}}}+\frac{q_{0}\lr{A}^2 V\lr{\lambda_0} f_0^{\prime}\lr{A}}{3 f_{0}\lr{A}^{3/2}}-\frac{f_{0}^{\prime}\lr{A}^2}{f_{0}\lr{A}^{3/2}}=0,\\
    &q_0\lr{A} \left(-\frac{f_{0}^{\prime}\lr{A}}{f_0\lr{A}}-4\right)+\frac{q_0\lr{A}^3 V\lr{\lambda_0}}{3 f_0\lr{A}}+q_{0}^{\prime}=0,\\
    &\frac{1}{2} \sqrt{\frac{9 f_{0}^{\prime}\lr{A}+36 f_0\lr{A}-3 q_{0}\lr{A}^2 V\lr{\lambda_0}}{f_{0}\lr{A}}}+\frac{\lambda_{0}^{\prime}\lr{A}}{\lambda_0\lr{A}}=0 \, ,
\end{split}
\end{equation}
subject to the following horizon boundary conditions
\begin{equation}
\label{horizon conditions}
\begin{split}
    &f_0=f_{00}\lr{A-A_h}+f_{01}\lr{A-A_h}^2+\dots \\
    &\lambda_0=\lambda_{00}+\lambda_{01}\lr{A-A_h}+\dots\\
    &q_0=q_{00}+q_{01}\lr{A-A_h}+\dots \, .
\end{split}
\end{equation}
Regularity at the horizon in these coordinates implies:
\begin{equation}
    q_{00}^2=\frac{3f_{00}}{V_h}.
\end{equation}
Notice that in this form the equations (\ref{numerical equations}) are invariant under a resealing of the type
\begin{equation}
\label{rescale}
    f_0\to\alpha f_0 \, , \quad q_0\to\sqrt{\alpha}q_0.
\end{equation}
We use this in the numerical solution to scale $f_0\to1$ at the UV boundary. Thus, the procedure we follow is:
\begin{itemize}
  \item Solve the system of equations numerically assuming the horizon behavior (\ref{horizon conditions}) with $f_{00}=1$. $f(A)$ will then go to some value $f_{UV}$ at the boundary.
  \item Rescale $f_0\to f_0/f_{UV}\, , q_0\to q_0/\sqrt{f_{UV}}$ using (\ref{rescale}), so that the rescaled $f_0$ goes to 1 in the UV.
\end{itemize}
Then the solution is correctly normalized, with a fixed definition of the boundary time. 

The next step in the bulk viscosity calculation is to solve numerically the equation of motion obeyed by $z_{0}$, which is obtained from \eqref{EZ} at leading order in the hydrodynamic expansion. $z_0$ follows an expansion in powers of $\beta$ similar to \eqref{expbg}   
\begin{equation}
z_0 = z_{00} +\beta z_{01}+\mathcal{O}(\beta^2) \, ,
\end{equation}
where the leading order term $z_{00}$ obeys
\begin{equation}
\label{z_00 equation}
\begin{split}
&z_{00}^{\prime\prime}+z_{00}^{\prime}\left(\frac{f_{0}^{\prime}}{f_{0}}-\frac{q_{0}^{\prime}}{q_{0}}+4\right)+\\
&z_{00}\frac{q_{0}^{2} \left(3 f_{0} \left(64 V(\lambda_0\right)+\left(16\lambda_{0}^{\prime}+9\lambda_{0}\right) V'(\lambda_0)+9 \lambda_{0}^{2} V''(\lambda_0)-16 V(\lambda_0) \left(q_{0}^{2} V(\lambda_0)-3 f_{0}^{\prime}\right)\right)}{72f_{0}^{2}} =0.
\end{split}
\end{equation}
$z_{00}$ is assumed to be regular at the horizon, where it admits an expansion of the form:
\begin{equation}
    z_{00}=c_0+c_1\lr{A-A_h}.
\end{equation}
Substituting this into the $z_{00}$ equation along with the horizon expansion of the background (\ref{horizon conditions}), regularity at the horizon is found to impose a relation between the coefficients $c_0$ and $c_1$
\begin{equation}
\label{z0 horizon condition}
    c_1=c_0\frac{9}{8}\lr{\frac{ V'(\Phi_h)^2}{ V(\Phi_h)^2}-\frac{V''(\Phi_h)}{8 V(\Phi_h)}} \, .
\end{equation}
In addition, we must require that the dilaton perturbation $\Psi$ is not sourced, since only the metric should be sourced to compute the bulk viscosity. From \eqref{defZ} and \eqref{H11UV}, this fixes the leading behavior of $z_{00}$ near the boundary to be 
\begin{equation}
\label{z0 boundary condition}
    z_{00}\big|_{UV} \underset{A\to\infty}{\sim} 1/2A \, .
\end{equation}
With the two boundary conditions \eqref{z0 horizon condition}-\eqref{z0 boundary condition}, the fluctuation equations admit a unique solution that can be found numerically. Since we use a shooting method from the horizon to solve the equations, implementing the UV condition \eqref{z0 boundary condition} is not as straightforward as \eqref{z0 horizon condition}. It can be done with the following procedure:
\begin{itemize}
  \item We solve the equation with the previously computed background solution, imposing (\ref{z0 horizon condition}) at the horizon, with $c_0=1$.
  \item We rescale the solution appropriately such that $z_{00}$ satisfies (\ref{z0 boundary condition}). Notice that (\ref{z_00 equation}) is invariant under the rescaling of $z_{00}$.
\end{itemize}
Then $z_{00}$ matches the required boundary conditions. The bulk viscosity then follows from 
\begin{equation}
    \frac{\zeta}{s}=\frac{8z_{00}\lr{A_h}^2}{27\pi}\, .
\end{equation}
Once we have the order $\mathcal{O}(\beta^0)$ solution, the extension of the above procedure to include the order $\mathcal{O}(\beta)$ corrections is straightforward. The method for fixing the normalization will then be applied to the background functions $\{f,q,\lambda\}$ and the perturbation $z_0$ including the ${\cal O}(\beta)$ corrections. 

\providecommand{\href}[2]{#2}\begingroup\raggedright\endgroup


\begin{thebibliography}{10}

\bibitem{Busza:2018rrf}
W.~Busza, K.~Rajagopal, and W.~van~der Schee, {\it {Heavy Ion Collisions: The Big Picture, and the Big Questions}},  {\em Ann. Rev. Nucl. Part. Sci.} {\bf 68} (2018) 339--376, [\href{http://arxiv.org/abs/1802.04801}{{\tt arXiv:1802.04801}}].

\bibitem{Meyer:2007ic}
H.~B. Meyer, {\it {A Calculation of the shear viscosity in SU(3) gluodynamics}},  {\em Phys. Rev. D} {\bf 76} (2007) 101701, [\href{http://arxiv.org/abs/0704.1801}{{\tt arXiv:0704.1801}}].

\bibitem{Meyer:2007dy}
H.~B. Meyer, {\it {A Calculation of the bulk viscosity in SU(3) gluodynamics}},  {\em Phys. Rev. Lett.} {\bf 100} (2008) 162001, [\href{http://arxiv.org/abs/0710.3717}{{\tt arXiv:0710.3717}}].

\bibitem{Nijs:2020ors}
G.~Nijs, W.~van~der Schee, U.~G\"ursoy, and R.~Snellings, {\it {Transverse Momentum Differential Global Analysis of Heavy-Ion Collisions}},  {\em Phys. Rev. Lett.} {\bf 126} (2021), no.~20 202301, [\href{http://arxiv.org/abs/2010.15130}{{\tt arXiv:2010.15130}}].

\bibitem{Nijs:2020roc}
G.~Nijs, W.~van~der Schee, U.~G\"ursoy, and R.~Snellings, {\it {Bayesian analysis of heavy ion collisions with the heavy ion computational framework Trajectum}},  {\em Phys. Rev. C} {\bf 103} (2021), no.~5 054909, [\href{http://arxiv.org/abs/2010.15134}{{\tt arXiv:2010.15134}}].

\bibitem{Nijs:2023yab}
G.~Nijs and W.~van~der Schee, {\it {A generalized hydrodynamizing initial stage for Heavy Ion Collisions}},  \href{http://arxiv.org/abs/2304.06191}{{\tt arXiv:2304.06191}}.

\bibitem{Giacalone:2023cet}
G.~Giacalone, G.~Nijs, and W.~van~der Schee, {\it {Determination of the Neutron Skin of Pb208 from Ultrarelativistic Nuclear Collisions}},  {\em Phys. Rev. Lett.} {\bf 131} (2023), no.~20 202302, [\href{http://arxiv.org/abs/2305.00015}{{\tt arXiv:2305.00015}}].

\bibitem{JETSCAPE:2020mzn}
{\bf JETSCAPE} Collaboration, D.~Everett et~al., {\it {Multisystem Bayesian constraints on the transport coefficients of QCD matter}},  {\em Phys. Rev. C} {\bf 103} (2021), no.~5 054904, [\href{http://arxiv.org/abs/2011.01430}{{\tt arXiv:2011.01430}}].

\bibitem{Denicol:2014vaa}
G.~S. Denicol, S.~Jeon, and C.~Gale, {\it {Transport Coefficients of Bulk Viscous Pressure in the 14-moment approximation}},  {\em Phys. Rev. C} {\bf 90} (2014), no.~2 024912, [\href{http://arxiv.org/abs/1403.0962}{{\tt arXiv:1403.0962}}].

\bibitem{Moreland:2014oya}
J.~S. Moreland, J.~E. Bernhard, and S.~A. Bass, {\it {Alternative ansatz to wounded nucleon and binary collision scaling in high-energy nuclear collisions}},  {\em Phys. Rev. C} {\bf 92} (2015), no.~1 011901, [\href{http://arxiv.org/abs/1412.4708}{{\tt arXiv:1412.4708}}].

\bibitem{Cooper:1974mv}
F.~Cooper and G.~Frye, {\it {Comment on the Single Particle Distribution in the Hydrodynamic and Statistical Thermodynamic Models of Multiparticle Production}},  {\em Phys. Rev. D} {\bf 10} (1974) 186.

\bibitem{Trachenko:2020ktm}
K.~Trachenko, V.~Brazhkin, and M.~Baggioli, {\it {Similarity between the kinematic viscosity of quark-gluon plasma and liquids at the viscosity minimum}},  {\em SciPost Phys.} {\bf 10} (2021), no.~5 118, [\href{http://arxiv.org/abs/2003.13506}{{\tt arXiv:2003.13506}}].

\bibitem{Cremonini:2012ny}
S.~Cremonini, U.~Gursoy, and P.~Szepietowski, {\it {On the Temperature Dependence of the Shear Viscosity and Holography}},  {\em JHEP} {\bf 08} (2012) 167, [\href{http://arxiv.org/abs/1206.3581}{{\tt arXiv:1206.3581}}].

\bibitem{Maldacena:1997re}
J.~M. Maldacena, {\it {The Large N limit of superconformal field theories and supergravity}},  {\em Adv. Theor. Math. Phys.} {\bf 2} (1998) 231--252, [\href{http://arxiv.org/abs/hep-th/9711200}{{\tt hep-th/9711200}}].

\bibitem{Gubser:1998bc}
S.~S. Gubser, I.~R. Klebanov, and A.~M. Polyakov, {\it {Gauge theory correlators from noncritical string theory}},  {\em Phys. Lett. B} {\bf 428} (1998) 105--114, [\href{http://arxiv.org/abs/hep-th/9802109}{{\tt hep-th/9802109}}].

\bibitem{Witten:1998qj}
E.~Witten, {\it {Anti-de Sitter space and holography}},  {\em Adv. Theor. Math. Phys.} {\bf 2} (1998) 253--291, [\href{http://arxiv.org/abs/hep-th/9802150}{{\tt hep-th/9802150}}].

\bibitem{Policastro:2001yc}
G.~Policastro, D.~T. Son, and A.~O. Starinets, {\it {The Shear viscosity of strongly coupled N=4 supersymmetric Yang-Mills plasma}},  {\em Phys. Rev. Lett.} {\bf 87} (2001) 081601, [\href{http://arxiv.org/abs/hep-th/0104066}{{\tt hep-th/0104066}}].

\bibitem{Gursoy:2007cb}
U.~Gursoy and E.~Kiritsis, {\it {Exploring improved holographic theories for QCD: Part I}},  {\em JHEP} {\bf 02} (2008) 032, [\href{http://arxiv.org/abs/0707.1324}{{\tt arXiv:0707.1324}}].

\bibitem{Gursoy:2007er}
U.~Gursoy, E.~Kiritsis, and F.~Nitti, {\it {Exploring improved holographic theories for QCD: Part II}},  {\em JHEP} {\bf 02} (2008) 019, [\href{http://arxiv.org/abs/0707.1349}{{\tt arXiv:0707.1349}}].

\bibitem{Jarvinen:2011qe}
M.~Jarvinen and E.~Kiritsis, {\it {Holographic Models for QCD in the Veneziano Limit}},  {\em JHEP} {\bf 03} (2012) 002, [\href{http://arxiv.org/abs/1112.1261}{{\tt arXiv:1112.1261}}].

\bibitem{Veneziano:1976wm}
G.~Veneziano, {\it {Some Aspects of a Unified Approach to Gauge, Dual and Gribov Theories}},  {\em Nucl. Phys. B} {\bf 117} (1976) 519--545.

\bibitem{Gursoy:2008bu}
U.~Gursoy, E.~Kiritsis, L.~Mazzanti, and F.~Nitti, {\it {Deconfinement and Gluon Plasma Dynamics in Improved Holographic QCD}},  {\em Phys. Rev. Lett.} {\bf 101} (2008) 181601, [\href{http://arxiv.org/abs/0804.0899}{{\tt arXiv:0804.0899}}].

\bibitem{Gursoy:2008za}
U.~Gursoy, E.~Kiritsis, L.~Mazzanti, and F.~Nitti, {\it {Holography and Thermodynamics of 5D Dilaton-gravity}},  {\em JHEP} {\bf 05} (2009) 033, [\href{http://arxiv.org/abs/0812.0792}{{\tt arXiv:0812.0792}}].

\bibitem{Gursoy:2009jd}
U.~Gursoy, E.~Kiritsis, L.~Mazzanti, and F.~Nitti, {\it {Improved Holographic Yang-Mills at Finite Temperature: Comparison with Data}},  {\em Nucl. Phys. B} {\bf 820} (2009) 148--177, [\href{http://arxiv.org/abs/0903.2859}{{\tt arXiv:0903.2859}}].

\bibitem{Alho:2012mh}
T.~Alho, M.~J\"arvinen, K.~Kajantie, E.~Kiritsis, and K.~Tuominen, {\it {On finite-temperature holographic QCD in the Veneziano limit}},  {\em JHEP} {\bf 01} (2013) 093, [\href{http://arxiv.org/abs/1210.4516}{{\tt arXiv:1210.4516}}].

\bibitem{Alho:2013hsa}
T.~Alho, M.~J\"arvinen, K.~Kajantie, E.~Kiritsis, C.~Rosen, and K.~Tuominen, {\it {A holographic model for QCD in the Veneziano limit at finite temperature and density}},  {\em JHEP} {\bf 04} (2014) 124, [\href{http://arxiv.org/abs/1312.5199}{{\tt arXiv:1312.5199}}]. [Erratum: JHEP 02, 033 (2015)].

\bibitem{Gursoy:2016ofp}
U.~G\"ursoy, I.~Iatrakis, M.~J\"arvinen, and G.~Nijs, {\it {Inverse Magnetic Catalysis from improved Holographic QCD in the Veneziano limit}},  {\em JHEP} {\bf 03} (2017) 053, [\href{http://arxiv.org/abs/1611.06339}{{\tt arXiv:1611.06339}}].

\bibitem{Gursoy:2017wzz}
U.~Gursoy, M.~Jarvinen, and G.~Nijs, {\it {Holographic QCD in the Veneziano Limit at a Finite Magnetic Field and Chemical Potential}},  {\em Phys. Rev. Lett.} {\bf 120} (2018), no.~24 242002, [\href{http://arxiv.org/abs/1707.00872}{{\tt arXiv:1707.00872}}].

\bibitem{Buchel:2003tz}
A.~Buchel and J.~T. Liu, {\it {Universality of the shear viscosity in supergravity}},  {\em Phys. Rev. Lett.} {\bf 93} (2004) 090602, [\href{http://arxiv.org/abs/hep-th/0311175}{{\tt hep-th/0311175}}].

\bibitem{Gubser:2008sz}
S.~S. Gubser, S.~S. Pufu, and F.~D. Rocha, {\it {Bulk viscosity of strongly coupled plasmas with holographic duals}},  {\em JHEP} {\bf 08} (2008) 085, [\href{http://arxiv.org/abs/0806.0407}{{\tt arXiv:0806.0407}}].

\bibitem{Gursoy:2009kk}
U.~Gursoy, E.~Kiritsis, G.~Michalogiorgakis, and F.~Nitti, {\it {Thermal Transport and Drag Force in Improved Holographic QCD}},  {\em JHEP} {\bf 12} (2009) 056, [\href{http://arxiv.org/abs/0906.1890}{{\tt arXiv:0906.1890}}].

\bibitem{Buchel:2007mf}
A.~Buchel, {\it {Bulk viscosity of gauge theory plasma at strong coupling}},  {\em Phys. Lett. B} {\bf 663} (2008) 286--289, [\href{http://arxiv.org/abs/0708.3459}{{\tt arXiv:0708.3459}}].

\bibitem{Buchel:2008uu}
A.~Buchel and C.~Pagnutti, {\it {Bulk viscosity of N=2* plasma}},  {\em Nucl. Phys. B} {\bf 816} (2009) 62--72, [\href{http://arxiv.org/abs/0812.3623}{{\tt arXiv:0812.3623}}].

\bibitem{Eling:2011ms}
C.~Eling and Y.~Oz, {\it {A Novel Formula for Bulk Viscosity from the Null Horizon Focusing Equation}},  {\em JHEP} {\bf 06} (2011) 007, [\href{http://arxiv.org/abs/1103.1657}{{\tt arXiv:1103.1657}}].

\bibitem{Buchel:2011wx}
A.~Buchel, U.~Gursoy, and E.~Kiritsis, {\it {Holographic bulk viscosity: GPR versus EO}},  {\em JHEP} {\bf 09} (2011) 095, [\href{http://arxiv.org/abs/1104.2058}{{\tt arXiv:1104.2058}}].

\bibitem{Buchel:2011uj}
A.~Buchel, {\it {Violation of the holographic bulk viscosity bound}},  {\em Phys. Rev. D} {\bf 85} (2012) 066004, [\href{http://arxiv.org/abs/1110.0063}{{\tt arXiv:1110.0063}}].

\bibitem{Ballon-Bayona:2021tzw}
A.~Ballon-Bayona, L.~A.~H. Mamani, A.~S. Miranda, and V.~T. Zanchin, {\it {Effective holographic models for QCD: Thermodynamics and viscosity coefficients}},  {\em Phys. Rev. D} {\bf 104} (2021), no.~4 046013, [\href{http://arxiv.org/abs/2103.14188}{{\tt arXiv:2103.14188}}].

\bibitem{Demircik:2023lsn}
T.~Demircik, D.~Gallegos, U.~G\"ursoy, M.~J\"arvinen, and R.~Lier, {\it {A Novel Method for Holographic Transport}},  \href{http://arxiv.org/abs/2311.00042}{{\tt arXiv:2311.00042}}.

\bibitem{Demircik:2024bxd}
T.~Demircik, D.~Gallegos, U.~G\"ursoy, M.~J\"arvinen, and R.~Lier, {\it {Holographic transport in anisotropic plasmas}},  {\em Phys. Rev. D} {\bf 110} (2024), no.~6 066007, [\href{http://arxiv.org/abs/2402.12224}{{\tt arXiv:2402.12224}}].

\bibitem{Buchel:2023fst}
A.~Buchel, S.~Cremonini, and L.~Early, {\it {Holographic transport beyond the supergravity approximation}},  {\em JHEP} {\bf 04} (2024) 032, [\href{http://arxiv.org/abs/2312.05377}{{\tt arXiv:2312.05377}}].

\bibitem{CruzRojas:2024etx}
J.~Cruz~Rojas, T.~Gorda, C.~Hoyos, N.~Jokela, M.~J\"arvinen, A.~Kurkela, R.~Paatelainen, S.~S\"appi, and A.~Vuorinen, {\it {Estimate for the Bulk Viscosity of Strongly Coupled Quark Matter Using Perturbative QCD and Holography}},  {\em Phys. Rev. Lett.} {\bf 133} (2024), no.~7 071901, [\href{http://arxiv.org/abs/2402.00621}{{\tt arXiv:2402.00621}}].

\bibitem{Kovtun:2003wp}
P.~Kovtun, D.~T. Son, and A.~O. Starinets, {\it {Holography and hydrodynamics: Diffusion on stretched horizons}},  {\em JHEP} {\bf 10} (2003) 064, [\href{http://arxiv.org/abs/hep-th/0309213}{{\tt hep-th/0309213}}].

\bibitem{Kovtun:2004de}
P.~Kovtun, D.~T. Son, and A.~O. Starinets, {\it {Viscosity in strongly interacting quantum field theories from black hole physics}},  {\em Phys. Rev. Lett.} {\bf 94} (2005) 111601, [\href{http://arxiv.org/abs/hep-th/0405231}{{\tt hep-th/0405231}}].

\bibitem{Kats:2007mq}
Y.~Kats and P.~Petrov, {\it {Effect of curvature squared corrections in AdS on the viscosity of the dual gauge theory}},  {\em JHEP} {\bf 01} (2009) 044, [\href{http://arxiv.org/abs/0712.0743}{{\tt arXiv:0712.0743}}].

\bibitem{Brigante:2007nu}
M.~Brigante, H.~Liu, R.~C. Myers, S.~Shenker, and S.~Yaida, {\it {Viscosity Bound Violation in Higher Derivative Gravity}},  {\em Phys. Rev. D} {\bf 77} (2008) 126006, [\href{http://arxiv.org/abs/0712.0805}{{\tt arXiv:0712.0805}}].

\bibitem{Myers:2008yi}
R.~C. Myers, M.~F. Paulos, and A.~Sinha, {\it {Quantum corrections to eta/s}},  {\em Phys. Rev. D} {\bf 79} (2009) 041901, [\href{http://arxiv.org/abs/0806.2156}{{\tt arXiv:0806.2156}}].

\bibitem{Cremonini:2011iq}
S.~Cremonini, {\it {The Shear Viscosity to Entropy Ratio: A Status Report}},  {\em Mod. Phys. Lett. B} {\bf 25} (2011) 1867--1888, [\href{http://arxiv.org/abs/1108.0677}{{\tt arXiv:1108.0677}}].

\bibitem{Grozdanov:2016zjj}
S.~Grozdanov and W.~van~der Schee, {\it {Coupling Constant Corrections in a Holographic Model of Heavy Ion Collisions}},  {\em Phys. Rev. Lett.} {\bf 119} (2017), no.~1 011601, [\href{http://arxiv.org/abs/1610.08976}{{\tt arXiv:1610.08976}}].

\bibitem{Folkestad:2019lam}
A.~Folkestad, S.~Grozdanov, K.~Rajagopal, and W.~van~der Schee, {\it {Coupling Constant Corrections in a Holographic Model of Heavy Ion Collisions with Nonzero Baryon Number Density}},  {\em JHEP} {\bf 12} (2019) 093, [\href{http://arxiv.org/abs/1907.13134}{{\tt arXiv:1907.13134}}].

\bibitem{Buchel:2010wf}
A.~Buchel and S.~Cremonini, {\it {Viscosity Bound and Causality in Superfluid Plasma}},  {\em JHEP} {\bf 10} (2010) 026, [\href{http://arxiv.org/abs/1007.2963}{{\tt arXiv:1007.2963}}].

\bibitem{Czajka:2018bod}
A.~Czajka, K.~Dasgupta, C.~Gale, S.~Jeon, A.~Misra, M.~Richard, and K.~Sil, {\it {Bulk Viscosity at Extreme Limits: From Kinetic Theory to Strings}},  {\em JHEP} {\bf 07} (2019) 145, [\href{http://arxiv.org/abs/1807.04713}{{\tt arXiv:1807.04713}}].

\bibitem{Yadav:2020tyo}
V.~Yadav and A.~Misra, {\it {On $M$-theory dual of large-$N$ thermal QCD-like theories up to $\mathcal{O}(R^4)$ and $G$-structure classification of underlying non-supersymmetric geometries}},  {\em Adv. Theor. Math. Phys.} {\bf 26} (2022), no.~10 3801--3894, [\href{http://arxiv.org/abs/2004.07259}{{\tt arXiv:2004.07259}}].

\bibitem{Kushwah:2024ngr}
S.~S. Kushwah and A.~Misra, {\it {Bulk viscosity, speed of sound, and contact structure at intermediate coupling}},  {\em Phys. Rev. D} {\bf 110} (2024), no.~12 126010, [\href{http://arxiv.org/abs/2403.10541}{{\tt arXiv:2403.10541}}].

\bibitem{Kiritsis:2009hu}
E.~Kiritsis, {\it {Dissecting the string theory dual of QCD}},  {\em Fortsch. Phys.} {\bf 57} (2009) 396--417, [\href{http://arxiv.org/abs/0901.1772}{{\tt arXiv:0901.1772}}].

\bibitem{Alho:2015zua}
T.~Alho, M.~Jarvinen, K.~Kajantie, E.~Kiritsis, and K.~Tuominen, {\it {Quantum and stringy corrections to the equation of state of holographic QCD matter and the nature of the chiral transition}},  {\em Phys. Rev. D} {\bf 91} (2015), no.~5 055017, [\href{http://arxiv.org/abs/1501.06379}{{\tt arXiv:1501.06379}}].

\bibitem{Gallegos}
D.~Gallegos, ``unpublished.''.

\bibitem{Hosoya:1983id}
A.~Hosoya, M.-a. Sakagami, and M.~Takao, {\it {Nonequilibrium Thermodynamics in Field Theory: Transport Coefficients}},  {\em Annals Phys.} {\bf 154} (1984) 229.

\bibitem{Son:2002sd}
D.~T. Son and A.~O. Starinets, {\it {Minkowski space correlators in AdS / CFT correspondence: Recipe and applications}},  {\em JHEP} {\bf 09} (2002) 042, [\href{http://arxiv.org/abs/hep-th/0205051}{{\tt hep-th/0205051}}].

\bibitem{Wald:1993nt}
R.~M. Wald, {\it {Black hole entropy is the Noether charge}},  {\em Phys. Rev. D} {\bf 48} (1993), no.~8 R3427--R3431, [\href{http://arxiv.org/abs/gr-qc/9307038}{{\tt gr-qc/9307038}}].

\bibitem{Shifman:1978bx}
M.~A. Shifman, A.~I. Vainshtein, and V.~I. Zakharov, {\it {QCD and Resonance Physics. Theoretical Foundations}},  {\em Nucl. Phys. B} {\bf 147} (1979) 385--447.

\bibitem{Novikov:1983jt}
V.~A. Novikov, M.~A. Shifman, A.~I. Vainshtein, M.~B. Voloshin, and V.~I. Zakharov, {\it {Use and Misuse of QCD Sum Rules, Factorization and Related Topics}},  {\em Nucl. Phys. B} {\bf 237} (1984) 525--552.

\bibitem{Novikov:1980uj}
V.~A. Novikov, M.~A. Shifman, A.~I. Vainshtein, and V.~I. Zakharov, {\it {OPERATOR EXPANSION IN QUANTUM CHROMODYNAMICS BEYOND PERTURBATION THEORY}},  {\em Nucl. Phys. B} {\bf 174} (1980) 378--396.

\bibitem{Jarvinen:2021jbd}
M.~J\"arvinen, {\it {Holographic modeling of nuclear matter and neutron stars}},  {\em Eur. Phys. J. C} {\bf 82} (2022), no.~4 282, [\href{http://arxiv.org/abs/2110.08281}{{\tt arXiv:2110.08281}}].

\bibitem{Panero:2009tv}
M.~Panero, {\it {Thermodynamics of the QCD plasma and the large-N limit}},  {\em Phys. Rev. Lett.} {\bf 103} (2009) 232001, [\href{http://arxiv.org/abs/0907.3719}{{\tt arXiv:0907.3719}}].

\bibitem{Demircik:2021zll}
T.~Demircik, C.~Ecker, and M.~J\"arvinen, {\it {Dense and Hot QCD at Strong Coupling}},  {\em Phys. Rev. X} {\bf 12} (2022), no.~4 041012, [\href{http://arxiv.org/abs/2112.12157}{{\tt arXiv:2112.12157}}].

\bibitem{Jokela:2021vwy}
N.~Jokela, M.~J\"arvinen, and J.~Remes, {\it {Holographic QCD in the NICER era}},  {\em Phys. Rev. D} {\bf 105} (2022), no.~8 086005, [\href{http://arxiv.org/abs/2111.12101}{{\tt arXiv:2111.12101}}].

\bibitem{Hoyos:2021njg}
C.~Hoyos, N.~Jokela, M.~J\"arvinen, J.~G. Subils, J.~Tarrio, and A.~Vuorinen, {\it {Holographic approach to transport in dense QCD matter}},  {\em Phys. Rev. D} {\bf 105} (2022), no.~6 066014, [\href{http://arxiv.org/abs/2109.12122}{{\tt arXiv:2109.12122}}].

\bibitem{Ecker:2019xrw}
C.~Ecker, M.~J\"arvinen, G.~Nijs, and W.~van~der Schee, {\it {Gravitational waves from holographic neutron star mergers}},  {\em Phys. Rev. D} {\bf 101} (2020), no.~10 103006, [\href{http://arxiv.org/abs/1908.03213}{{\tt arXiv:1908.03213}}].

\bibitem{Jokela:2018ers}
N.~Jokela, M.~J\"arvinen, and J.~Remes, {\it {Holographic QCD in the Veneziano limit and neutron stars}},  {\em JHEP} {\bf 03} (2019) 041, [\href{http://arxiv.org/abs/1809.07770}{{\tt arXiv:1809.07770}}].

\bibitem{Arean:2012mq}
D.~Arean, I.~Iatrakis, M.~J\"arvinen, and E.~Kiritsis, {\it {V-QCD: Spectra, the dilaton and the S-parameter}},  {\em Phys. Lett. B} {\bf 720} (2013) 219--223, [\href{http://arxiv.org/abs/1211.6125}{{\tt arXiv:1211.6125}}].

\bibitem{Buchel:2009tt}
A.~Buchel and R.~C. Myers, {\it {Causality of Holographic Hydrodynamics}},  {\em JHEP} {\bf 08} (2009) 016, [\href{http://arxiv.org/abs/0906.2922}{{\tt arXiv:0906.2922}}].

\bibitem{Brigante:2008gz}
M.~Brigante, H.~Liu, R.~C. Myers, S.~Shenker, and S.~Yaida, {\it {The Viscosity Bound and Causality Violation}},  {\em Phys. Rev. Lett.} {\bf 100} (2008) 191601, [\href{http://arxiv.org/abs/0802.3318}{{\tt arXiv:0802.3318}}].

\bibitem{Hohm:2010jy}
O.~Hohm, C.~Hull, and B.~Zwiebach, {\it {Background independent action for double field theory}},  {\em JHEP} {\bf 07} (2010) 016, [\href{http://arxiv.org/abs/1003.5027}{{\tt arXiv:1003.5027}}].

\bibitem{Hohm:2010pp}
O.~Hohm, C.~Hull, and B.~Zwiebach, {\it {Generalized metric formulation of double field theory}},  {\em JHEP} {\bf 08} (2010) 008, [\href{http://arxiv.org/abs/1006.4823}{{\tt arXiv:1006.4823}}].

\bibitem{Vafa:2005ui}
C.~Vafa, {\it {The String landscape and the swampland}},  \href{http://arxiv.org/abs/hep-th/0509212}{{\tt hep-th/0509212}}.

\bibitem{Ooguri:2006in}
H.~Ooguri and C.~Vafa, {\it {On the Geometry of the String Landscape and the Swampland}},  {\em Nucl. Phys. B} {\bf 766} (2007) 21--33, [\href{http://arxiv.org/abs/hep-th/0605264}{{\tt hep-th/0605264}}].

\bibitem{Palti:2019pca}
E.~Palti, {\it {The Swampland: Introduction and Review}},  {\em Fortsch. Phys.} {\bf 67} (2019), no.~6 1900037, [\href{http://arxiv.org/abs/1903.06239}{{\tt arXiv:1903.06239}}].

\bibitem{Kharzeev:2007jp}
D.~E. Kharzeev, L.~D. McLerran, and H.~J. Warringa, {\it {The Effects of topological charge change in heavy ion collisions: 'Event by event P and CP violation'}},  {\em Nucl. Phys. A} {\bf 803} (2008) 227--253, [\href{http://arxiv.org/abs/0711.0950}{{\tt arXiv:0711.0950}}].

\bibitem{Skokov:2009qp}
V.~Skokov, A.~Y. Illarionov, and V.~Toneev, {\it {Estimate of the magnetic field strength in heavy-ion collisions}},  {\em Int. J. Mod. Phys. A} {\bf 24} (2009) 5925--5932, [\href{http://arxiv.org/abs/0907.1396}{{\tt arXiv:0907.1396}}].

\bibitem{Tuchin:2013ie}
K.~Tuchin, {\it {Particle production in strong electromagnetic fields in relativistic heavy-ion collisions}},  {\em Adv. High Energy Phys.} {\bf 2013} (2013) 490495, [\href{http://arxiv.org/abs/1301.0099}{{\tt arXiv:1301.0099}}].

\bibitem{Voronyuk:2011jd}
V.~Voronyuk, V.~D. Toneev, W.~Cassing, E.~L. Bratkovskaya, V.~P. Konchakovski, and S.~A. Voloshin, {\it {(Electro-)Magnetic field evolution in relativistic heavy-ion collisions}},  {\em Phys. Rev. C} {\bf 83} (2011) 054911, [\href{http://arxiv.org/abs/1103.4239}{{\tt arXiv:1103.4239}}].

\bibitem{Deng:2012pc}
W.-T. Deng and X.-G. Huang, {\it {Event-by-event generation of electromagnetic fields in heavy-ion collisions}},  {\em Phys. Rev. C} {\bf 85} (2012) 044907, [\href{http://arxiv.org/abs/1201.5108}{{\tt arXiv:1201.5108}}].

\bibitem{Tuchin:2013apa}
K.~Tuchin, {\it {Time and space dependence of the electromagnetic field in relativistic heavy-ion collisions}},  {\em Phys. Rev. C} {\bf 88} (2013), no.~2 024911, [\href{http://arxiv.org/abs/1305.5806}{{\tt arXiv:1305.5806}}].

\bibitem{McLerran:2013hla}
L.~McLerran and V.~Skokov, {\it {Comments About the Electromagnetic Field in Heavy-Ion Collisions}},  {\em Nucl. Phys. A} {\bf 929} (2014) 184--190, [\href{http://arxiv.org/abs/1305.0774}{{\tt arXiv:1305.0774}}].

\bibitem{Gursoy:2014aka}
U.~Gursoy, D.~Kharzeev, and K.~Rajagopal, {\it {Magnetohydrodynamics, charged currents and directed flow in heavy ion collisions}},  {\em Phys. Rev. C} {\bf 89} (2014), no.~5 054905, [\href{http://arxiv.org/abs/1401.3805}{{\tt arXiv:1401.3805}}].

\bibitem{Gursoy:2018yai}
U.~G\"ursoy, D.~Kharzeev, E.~Marcus, K.~Rajagopal, and C.~Shen, {\it {Charge-dependent Flow Induced by Magnetic and Electric Fields in Heavy Ion Collisions}},  {\em Phys. Rev. C} {\bf 98} (2018), no.~5 055201, [\href{http://arxiv.org/abs/1806.05288}{{\tt arXiv:1806.05288}}].

\bibitem{Kharzeev:2013jha}
D.~Kharzeev, K.~Landsteiner, A.~Schmitt, and H.-U. Yee, eds., {\em {Strongly Interacting Matter in Magnetic Fields}}, vol.~871.
\newblock 2013.

\bibitem{Iqbal:2008by}
N.~Iqbal and H.~Liu, {\it {Universality of the hydrodynamic limit in AdS/CFT and the membrane paradigm}},  {\em Phys. Rev. D} {\bf 79} (2009) 025023, [\href{http://arxiv.org/abs/0809.3808}{{\tt arXiv:0809.3808}}].

\bibitem{DHoker:2009mmn}
E.~D'Hoker and P.~Kraus, {\it {Magnetic Brane Solutions in AdS}},  {\em JHEP} {\bf 10} (2009) 088, [\href{http://arxiv.org/abs/0908.3875}{{\tt arXiv:0908.3875}}].

\bibitem{Ballon-Bayona:2013cta}
A.~Ballon-Bayona, {\it {Holographic deconfinement transition in the presence of a magnetic field}},  {\em JHEP} {\bf 11} (2013) 168, [\href{http://arxiv.org/abs/1307.6498}{{\tt arXiv:1307.6498}}].

\bibitem{Rougemont:2015oea}
R.~Rougemont, R.~Critelli, and J.~Noronha, {\it {Holographic calculation of the QCD crossover temperature in a magnetic field}},  {\em Phys. Rev. D} {\bf 93} (2016), no.~4 045013, [\href{http://arxiv.org/abs/1505.07894}{{\tt arXiv:1505.07894}}].

\bibitem{Finazzo:2016mhm}
S.~I. Finazzo, R.~Critelli, R.~Rougemont, and J.~Noronha, {\it {Momentum transport in strongly coupled anisotropic plasmas in the presence of strong magnetic fields}},  {\em Phys. Rev. D} {\bf 94} (2016), no.~5 054020, [\href{http://arxiv.org/abs/1605.06061}{{\tt arXiv:1605.06061}}]. [Erratum: Phys.Rev.D 96, 019903 (2017)].

\bibitem{Gursoy:2020kjd}
U.~G\"ursoy, M.~J\"arvinen, G.~Nijs, and J.~F. Pedraza, {\it {On the interplay between magnetic field and anisotropy in holographic QCD}},  {\em JHEP} {\bf 03} (2021) 180, [\href{http://arxiv.org/abs/2011.09474}{{\tt arXiv:2011.09474}}].

\bibitem{Gursoy:2021efc}
U.~Gursoy, {\it {Holographic QCD and magnetic fields}},  {\em Eur. Phys. J. A} {\bf 57} (2021), no.~7 247, [\href{http://arxiv.org/abs/2104.02839}{{\tt arXiv:2104.02839}}].

\bibitem{Buchel:2004di}
A.~Buchel, J.~T. Liu, and A.~O. Starinets, {\it {Coupling constant dependence of the shear viscosity in N=4 supersymmetric Yang-Mills theory}},  {\em Nucl. Phys. B} {\bf 707} (2005) 56--68, [\href{http://arxiv.org/abs/hep-th/0406264}{{\tt hep-th/0406264}}].

\bibitem{Myers:2009ij}
R.~C. Myers, M.~F. Paulos, and A.~Sinha, {\it {Holographic Hydrodynamics with a Chemical Potential}},  {\em JHEP} {\bf 06} (2009) 006, [\href{http://arxiv.org/abs/0903.2834}{{\tt arXiv:0903.2834}}].

\bibitem{Cremonini:2011ej}
S.~Cremonini and P.~Szepietowski, {\it {Generating Temperature Flow for eta/s with Higher Derivatives: From Lifshitz to AdS}},  {\em JHEP} {\bf 02} (2012) 038, [\href{http://arxiv.org/abs/1111.5623}{{\tt arXiv:1111.5623}}].

\end{thebibliography}
\end{document}